\documentclass[paper]{geophysics}

\usepackage[tableposition=top]{caption}
\usepackage [english]{babel}
\usepackage [autostyle, english = american]{csquotes}
\usepackage{amsmath}
\usepackage{algorithm2e}
\usepackage{algpseudocode}
\usepackage{cleveref}
\MakeOuterQuote{"}
\usepackage{array}
\usepackage{makecell}
\usepackage{tabularx,booktabs}
\newcolumntype{Y}{>{\centering\arraybackslash}X}

\newcommand\captionOne{BGBM workflow after \cite{Pradhan2020_ABC}.}
\newcommand\captionTwo{Possible procedures to estimate the seismic data likelihood for BGBM workflow. Scheme of three distinct labeled kinematic models to calculate proxies as a metric for seismic data likelihood when utilizing the computed velocity of a sample.}
\newcommand\captionThree{Benchmark proxy values of samples after well conditioning used in the real case application. The dashed orange line represents the threshold value computed from the legacy velocity for the ABC rejection sampling strategy. Two antagonistic realizations were selected to illustrate the general ADCIGs behavior of seismic data with the velocity derived from the respective realization.}
\newcommand\captionFour{Semblance proxy values of samples after well data conditioning used in the real case application. The dashed orange line represents the threshold value computed from the legacy velocity. Two antagonistic realizations were selected to illustrate the general behavior of their RMS velocity in the semblance velocity spectra.}
\newcommand\captionFive{Two-dimensional MDS representation of resultant seismic horizons depth-positioning of prior samples used in the real case application. Horizons proxy values are illustrated by color, and the dashed orange circle represents the threshold value computed from the heuristic assumption. Two opposite samples, with rim colors connecting their representation as horizons, were selected to describe the general behavior of their resultant seismic horizons depth-positioning after the depth-time and time-depth conversions. The full red lines in horizons at depth plots represent the original depth-positioning of seismic horizons.}
\newcommand\captionSix{(a) The depth-migrated seismic image used in the real case study overlain by the interpreted stratigraphic horizons. (b) Corresponding structural input for the basin modeling simulations. The red lines represent the interpreted horizons, and the black lines are the interface of sub-layers considered in the basin modeling. Each capital letter represents the geological epoch/age of the referred stack of layers. The magenta vertical line represents the location of Well-1, and the vertical golden line illustrates the projection of the blind well used as a validation test.}
\newcommand\captionSeven{The transformation of each Monte Carlo realization from basin modeling sub-surface fields results in an associated velocity field through the established rock physics model.}
\newcommand\captionEight{Steps for pre-posterior distributions. (a) Illustration of step 1: vertical profiles of modeled present-day porosity, pore pressure, and temperature at well-1. Grey curves represent prior realizations, and black curves come from the 30 selected models with the highest well data likelihood. Black dots with white rims depict the measured data. (b) Illustration of step 2: Prior (gray) and pre-posterior (black) distributions of basin modeling input parameters: layer 07 uncertain parameters and heat flow value.}
\newcommand\captionNine{Summary of the IS strategy shown in horizons proxy MDS reduced space. The color represents the horizons proxy values, and the red diamond point is the projection of the original depth-positioning of seismic horizons in this space. (a) 2,500 prior samples; (b) 30 selected samples with the lowest well-1 data mismatch - samples used to estimate the pre-posterior distributions for importance sampling; (c) the 1,000 realizations with the lowest well-1 data mismatch from the 10,000 drawn samples from pre-posterior distributions IS resampling - subsequent seismic data likelihood estimation will be performed on these 1,000 samples. The numbers below the plots are associated with the steps illustrated in Figure \ref{fig:Figs/Figure8}.}
\newcommand\captionTen{The posterior selected samples for each seismic data likelihood proxy highlighted in the reduced MDS space of resultant depth-positioning of seismic horizons. The red diamond point is the projection of the original depth-positioning of seismic horizons in this space.}
\newcommand\captionEleven{(a) Vertical pore pressure profiles (gray lines) of priors and posteriors (colored lines according to proxy used for seismic data likelihood) at the projection of the blind well on the 2D studied section. The red dot represents the blind well mudweight measurement, and the black lines are the pore pressure boundaries: hydrostatic pressure on the left and vertical stress on the right. (b) The marginal probability density functions computed by KDE for the pore pressure prediction at mudweight depth measurement $z=3331.0m$.}
\newcommand\captionTwelve{Cross-plot between the benchmark proxy values with the statistical distance considering transformed velocity fields for all 1,000 samples used to evaluate the seismic data likelihood ($f(\mathbf{d}_{\mathbf{s},\mathbf{obs}}\left|\mathbf{h},\mathbf{b}\right)$). Types of velocity used in the experiment: (a) Interval velocity, (b) slowness, (c) time-domain RMS velocity, and (d) depth-averaged velocity. The horizontal dashed line represents the benchmark proxy threshold and is added as a reference to identify posterior samples.}
\newcommand\captionThirteen{Statistical percentiles (gray lines) of posterior ensembles of resultant depth-positioning of horizons and the original depth-positioning of horizons (black line). The gray dashed lines represent the 10\% and 90\% percentiles, whereas the solid gray one depicts the median. The posterior ensembles were retrieved based on seismic data likelihood proxies: (a) Benchmark; (b) Semblance.}
\newcommand\captionFourteen{Semblance velocity spectrum by NMO computation of two common mid-point locations of the studied 2D section. The black dashed lines represent the RMS velocity obtained from the legacy seismic velocity crossing the semblance scan.}
\newcommand\captionFifteen{Comparison of posterior results when using the horizons proxy with and without weights following depth values. (a) The marginal probability density functions for pore pressure prediction at the mudweight depth of blind well using posteriors retrieved from horizons proxy experiments. The other two sub-figures illustrate the statistical percentiles of horizons proxy posteriors ensemble of resultant depth-positioning of horizons by gray lines, and the original depth-positioning of horizons by black lines: (b) proxy computed with weights following depth - the horizons proxy used in this work, (c) constant weights considered for proxy computation - regular standardization.}

\newcommand\captionAOne{Prior falsification: (a) Vertical prior profiles in gray lines at the well-1 location of measured properties, whereas the well data is represented as dots; (b) Statistical robustness of prior resultant depth-positioning of horizons by Robust Mahalanobis Distance (RMD) in the six principal component scores retrieved from dimension reduction. The dashed line depicts the considered threshold for falsification.}

\newcommand\tcaptionOne{Summary of seismic data likelihood proxies}
\newcommand\tcaptionAOne{Prior distributions $N(\mu,\sigma)$ of the uncertain lithological parameters}
\newcommand\tcaptionATwo{Upper and lower limits of the prior distributions shown in Table \ref{table:priors}. The limits are fixed within each parameter type}
\newcommand\tcaptionAThree{Calibrated end-member parameters for rock physics model}

\begin{document}

\title{Bayesian Geophysical Basin Modeling with Seismic Kinematics Metrics to Quantify Uncertainty for Pore Pressure Prediction}
\renewcommand{\thefootnote}{\fnsymbol{footnote}} 

\address{
\footnotemark[1]Stanford University, Department of Energy Sciences \& Engineering, Stanford, California 94305-2220, USA. E-mail: josuesdf@stanford.edu (corresponding
author), pradhan1@stanford.edu, mukerji@stanford.edu

\footnotemark[2]California Institute of Technology, Pasadena, California, USA

\footnotemark[3]Stanford University, Department of Geophysics, Stanford, California 94305, USA

\footnotemark[4]Stanford University, Department of Earth \& Planetary Sciences, Stanford, California 94305, USA
}
\author{Josue Fonseca\footnotemark[1], Anshuman Pradhan\footnotemark[1]\footnotemark[2], Tapan Mukerji\footnotemark[1]\footnotemark[3]\footnotemark[4]}

\begin{abstract}
    Bayesian geophysical basin modeling (BGBM) methodology is an interdisciplinary workflow that incorporates data, geological expertise, and physical processes through Bayesian inference in sedimentary basin models. Its application culminates in subsurface models and properties that integrate the geo-history of a basin, rock physics definitions, well log and drilling data, and seismic information. Monte Carlo basin modeling realizations are performed by sampling from prior probability distributions on facies parameters and basin boundary conditions. After data assimilation, the accepted set of posterior sub-surface models yields uncertainty quantification of subsurface properties. This procedure is especially suitable for pore pressure prediction in a predrill stage. However, the high computational cost of seismic data assimilation decreases the practicality of the workflow. Therefore, we introduce and investigate seismic traveltimes criteria as computationally faster proxies for analyzing the seismic data likelihood when employing BGBM. The proposed surrogate schemes weigh the prior basin model results with the available seismic data with no need to perform expensive seismic depth-migration procedures for each Monte Carlo realization. Furthermore, we apply BGBM in a real field case from the Gulf of Mexico using a 2D section for pore pressure prediction considering different kinematics criteria. The workflow implementation with the novel seismic data assimilation proxies is compared with the complete computationally expensive benchmark approach, which utilizes a global analysis of the residual moveout in depth-migrated seismic image samples. Moreover, we validate and compare the outcomes for predicted pore pressure with mud-weight data from a blind well. The fast proxy of analyzing the depth-positioning of seismic horizons proposed in this work yields similar uncertainty quantification results in pore pressure prediction compared to the computationally expensive benchmark. The proposed fast proxies make the BGBM methodology efficient and practical. 
\end{abstract}


\footer{Submitted}
\lefthead{Fonseca}
\righthead{\emph{Fast BGBM for Pore Pressure Prediction}}


\section{Introduction}

Meaningful information about subsurface properties is essential in reducing economic risks and potential hazards in projects related to pore fluids in rocks. Therefore, a robust forecast of rock and fluid properties, especially pore pressure, supports environmental and human life protection before and during drilling projects. Thus, geoscientists are constantly investigating how to effectively improve the assessment of Earth models from geophysical data and geological knowledge.

Overpressure detection is most valuable before drilling. Thus, estimated velocities from reflection seismic methods are used to predict pore pressure away from wells on the sub-surface. These predrill procedures exploit the observation that overpressured intervals often have lower velocities than normally pressured ones at the same depth \cite[]{Mukerji2002_CSEG}. Nevertheless, pore pressure evaluation from seismic velocities generally adopts empirical methodologies \cite[]{Sayers02, Doyen2003} that may not be consistent with the geo-history of the basin.

Reliable Earth models should be geologically consistent and constrained by bounds derived from our physical understanding of the rocks. \cite{Peng2007} and \cite{Petmecky2009} adopted basin modeling results to derive geologically consistent velocities in order to enhance sub-salt seismic images with re-migration procedures. In addition, basin modeling provides a quantitative framework to simulate sub-surface properties coherent with physical processes occurring throughout the geo-history \cite[]{BPSM_book}.

\cite{Brevik_2014} coined the term geophysical basin modeling (GBM) as the procedure of estimating velocity models by integrating rock physics with basin model outcomes. The geo-history of porosity, effective stress, temperature and mineral volume fractions ultimately affect the velocity of rocks perceived by seismic waves. Consequently, basin model simulations provide the sub-surface properties required as input for the generation of geologically-consistent velocities through rock physics models. Thus, the authors advocate the adoption of a geologically derived velocity as an initial model to estimate seismic velocity through optimization methods. Then, one can extract enhanced seismic images when re-migrating with the latter result.

\cite{Szydlik2015} illustrate a GBM application in a geologically complex area located in deepwater Gulf of Mexico (GoM), showing significant improvement in sub-salt images. In addition to consistency with the basin geo-history, it is critical to quantify the uncertainties involved in Earth models to access the information given by subsurface data robustly. \cite{DePrisco2015} partially captured some of the uncertainties by employing GBM with several geologic scenarios. Through Monte Carlo sampling, the authors exhibit uncertainties related to velocity models through basin modeling results matching the well data.

The aforementioned works have focused on seismic velocity estimation to enhance seismic images when applying the GBM methodology. \cite{Pradhan2020_ABC} introduced a systematic workflow for uncertainty quantification when employing GBM along with Bayesian inference. Instead of using basin modeling only for velocities to improve seismic images from re-migrations, these authors put the basin modeling outcomes in a Bayesian framework as multiple possible posterior Earth models after the assimilation of well log data and seismic data. The data assimilation occurs by accepting from the prior Monte Carlo basin modeling realizations only a subset of posterior models that agree with available data from wells and seismic surveys. Therefore, one can retrieve a set of possible Earth models honoring geological expertise and physical processes conditioned to well and seismic data. Moreover, the set of posterior outcomes of GBM in a Bayesian framework yields uncertainty quantification for sub-surface properties. This practice is especially relevant for pore pressure prediction. The procedure by incorporating basin modeling allows the crucial and robust inclusion of possible overpressure generation mechanisms \cite[]{Osborne_AAPG97}. We call the methodology of utilizing a Bayesian inference with GBM as Bayesian Geophysical Basin Modeling (BGBM).

However, the BGBM workflow entails high computational costs as proposed by \cite{Pradhan2020_ABC}. Each Monte Carlo basin model sample undergoes a seismic depth-migration to analyze its seismic data likelihood. As one deals with thousands of Monte Carlo realizations, the number of seismic migration procedures is cumbersome even when evaluating a 2D section. In order to increase the practicability and efficiency of BGBM workflow, this paper focuses on reducing the computational time when assessing whether a basin model output realization is consistent with the measured seismic data. We propose using novel proxies to efficiently compute the seismic data likelihood of a realization rather than processing seismic depth-migrations for the proxy described by Ibid.

All these proxies explore expected kinematic properties from the seismic data along with the velocity obtained through rock physics modeling on the outcomes of each basin modeling realization. Depending on the kinematic property analyzed to estimate the seismic data likelihood of a sample, the computational demand and the way of handling the seismic data changes. The two proposed proxies not based on seismic depth-migrations drastically reduce the requirement for computational resources. Moreover, these new metrics utilize the seismic data in states more manageable and practical for day-to-day operations. These suggested proxies are based on: (a) the semblance evaluation in the normal moveout space; and (b) the analysis of statistical distances between (b.1) the depth-position of converted horizons for each velocity computed from realizations with (b.2) the original depth-positioning of input horizons used in the basin model interpreted on post-stack seismic images. Note that time-migration velocity spectra and post-stack seismic images are easier to handle than unmigrated pre-stack seismic data.

In summation, the intended contribution of this paper is to establish a fast BGBM implementation making it practical. The BGBM framework rigorously propagates prior geologic uncertainty to generate geologically consistent and physically valid Earth models with the assimilation of well and seismic data. Furthermore, utilizing the proposed surrogate kinematic models yields computational time reduction when estimating posterior realizations. Moreover, the new proxies retrieve kinematic seismic information from functional states of seismic data. Therefore, BGBM becomes a practical workflow for uncertainty quantification of sub-surface properties considering the novel proxies for seismic data assimilation.

The rest of the paper is organized into four sections: Methodology, Real case application, Discussion, and Conclusion. First, we provide a detailed treatment of the fast BGBM workflow and a description of the three possible proxies for calibrating the results with seismic data. The first and benchmark proxy (computationally expensive) was established by \cite{Pradhan2020_ABC}. The other two proxies (computationally fast) are the novel ones proposed in this work. Furthermore, we apply the BGBM methodology with all three proxies on a real case using a 2D section of the northern Gulf of Mexico basin. We utilize blind well data to validate the resulting uncertainty quantification of pore pressure prediction for all BGBM proxy cases. Afterwards, we discuss the nuances when selecting a particular proxy for the seismic data likelihood computation and compare the results obtained with them. Finally, we present the advantages and limitations of each case, along with future research directions.

\section{Methodology}

Bayesian geophysical basin modeling (BGBM) is based on the Popper-Bayes philosophy \cite[]{Tarantola2006}. This ideology suggests tackling inverse problems in a Bayesian framework along with falsification in order to define useful models through measured data. Therefore, one may establish a set of non-falsified Earth models by well data and a kinematics criterion for seismic data. These Earth models emanate from geologically derived prior distributions that are inputs to generative forward models (i.e., basin models). The generative models emulate geological and physical processes within a sedimentary basin and incorporate rock physics principles.  

\cite{Pradhan2020_ABC} were the first to introduce the BGBM idea. However, their workflow considers a computationally expensive criterion for evaluating the seismic kinematics associated with each Monte Carlo basin model realization. This paper introduces other possible computationally faster metrics to appraise the kinematic properties associated with the Monte Carlo samples. This section details how to perform BGBM and the reasoning behind the workflow. Moreover, we describe how to apply all the kinematics criteria mentioned in this work that may serve as a proxy for the seismic data likelihood of Monte Carlo realizations.	
    
\subsection{BGBM workflow}	

    The BGBM workflow incorporates geological expertise by establishing prior distributions of uncertain basin modeling input parameters. Then, Monte Carlo forward simulations of basin models incorporate physical processes throughout geological time. Afterwards, the basin modeling outputs such as porosity, mineralogy, temperature, and effective stress are used as inputs to rock physics models for elastic moduli to give a realistic, geologically-consistent velocity. Finally, the prior Earth models that are not falsified against available well data and seismic data are retained as posterior samples. 

    Figure \ref{fig:Figs/Figure1} illustrates the procedures and assumptions to implement the BGBM workflow. In the following, we describe the notation. The required forward modeling procedures are denoted as a function $g\left(.\right)$ with appropriate subscripts: basin modeling ($g_{BM}\left(.\right)$) rock physics modeling ($g_{RPM}\left(.\right)$); and seismic kinematic modeling ($g_{KM}\left(.\right)$). The probability density functions of random vectors are represented by $f\left(.\right)$. The available data ($\mathbf{d}_{\mathbf{obs}}$) is partitioned into data from well ($\mathbf{d}_{\mathbf{w},\mathbf{obs}}$) and seismic data ($\mathbf{d}_{s,\mathbf{obs}}$). These sub-elements are considered orthogonal to each other, i.e., conditionally independent given the Earth model. The random vector $\mathbf{b}$ describes the set of geological properties used as inputs for basin modeling (model parameters drawn from prior distributions). The random vector $\mathbf{h}$ represents the fields (2D or 3D) of sub-surface properties (e.g., porosity, temperature, pore pressure, velocity, etc.) generated through forward basin modeling and rock physics modeling.

    \plot{Figs/Figure1}{width=\columnwidth}{\captionOne}

    Moreover, $\mathbf{h}$ is partitioned into the fields obtained from basin modeling results ($\mathbf{h}_{-\mathbf{v}}=g_{BM}\left(\mathbf{b}\right)$) and the P-wave velocity field ($\mathbf{h}_\mathbf{v}$) derived by rock physics modeling. Porosity, vertical stress (overburden), pore pressure, temperature, and mineral distribution are examples from $\mathbf{h}_{-\mathbf{v}}$. Furthermore, the velocity field $\mathbf{h}_\mathbf{v}$ is derived using the simulated fields $\mathbf{h}_{-\mathbf{v}}$ as inputs to the rock physics model, i.e., $\mathbf{h}_\mathbf{v}=g_{RPM}\left(\mathbf{h}_{-\mathbf{v}}\right)$. One may note that any arbitrary randomly drawn $\mathbf{h}$ may include non-physical and geologically unfeasible samples. However, these points are avoided through the BGBM application, as only physically and geologically plausible $\mathbf{h}$’s are computed from the basin modeling. When the sub-surface properties ($\mathbf{h}$) are computed from samples $\mathbf{b}$, only a subset of possible $\mathbf{h}$ is accessible. This subset of $\mathbf{h}$ is called as $\mathbf{h}_{geologic}\equiv\mathbf{h}_\mathbf{g}$.
    
    No uncertainty in the forward simulation of models (apart from the input parameter uncertainty) is assumed in this study. Therefore, each Monte Carlo sample from the prior distribution of inputs to the forward simulation generates deterministic sub-surface property fields. The underlying premise, mathematically speaking, is that a $\mathbf{b}$ sample only derives a particular $\mathbf{h}$ through a function. This function exclusively yields points belonging to $\mathbf{h}_\mathbf{g}$ space. Since the Dirac delta function ($\delta\left(.\right)$) represents deterministic results in a probabilistic framework, the prior joint distribution of $\mathbf{h}$ and $\mathbf{b}$ is described in probability functions as follows:
	
    \begin{equation}
    \label{eq:unfold1_joint}
        f(\mathbf{h},\mathbf{b}) 
        = f(\mathbf{h} \mid \mathbf{b}) f(\mathbf{b})
        = f(\mathbf{h}_{\mathbf{v}} \mid \mathbf{h}_{-\mathbf{v}}, \mathbf{b}) f(\mathbf{h}_{-\mathbf{v}} \mid \mathbf{b}) f(\mathbf{b})
    \end{equation}
    \begin{equation}
    \label{eq:unfold2_joint}
        f(\mathbf{h},\mathbf{b})
        = \delta(\mathbf{h}_\mathbf{v} - g_{RPM}(\mathbf{h}_{-\mathbf{v}}))
        \delta(\mathbf{h}_{-\mathbf{v}} - g_{BM}(\mathbf{b})) f(\mathbf{b})
    \end{equation}

    Finally, the computed prior samples $\mathbf{h}_\mathbf{g}$ must be conditioned to available data ($\mathbf{d}_{\mathbf{obs}}$). Therefore, when conditioning to informative data, one expects a narrower posterior conditional distribution than the prior $\mathbf{h}_\mathbf{g}$. Considering the assumption of orthogonality (conditional independence) between the types of data, one may represent the posterior distribution obtained with BGBM as follows (with $c_0$ being a normalization constant):
	    
    \begin{equation}
    \label{eq:Bayes_inference}
        f(\mathbf{h}_{\mathbf{g}} \mid \mathbf{d}_{\mathbf{obs}}) 
        = c_0 f(\mathbf{d}_{\mathbf{obs}} \mid \mathbf{h}_{\mathbf{g}}) f(\mathbf{h}_{\mathbf{g}}) \\
        = c_0 f(\mathbf{d}_{\mathbf{w},\mathbf{obs}} \mid \mathbf{h}_{\mathbf{-v}}, \mathbf{b}) f(\mathbf{d}_{\mathbf{s},\mathbf{obs}} \mid \mathbf{h}_{v}, \mathbf{b}) f(\mathbf{h},\mathbf{b}) 
    \end{equation}

    The conditional independence assumption between the well data and seismic data allows us to divide the likelihood into two multiplicative terms. Thus, simulated sub-surface fields not directly associated with seismic data, such as porosity, pore pressure, and temperature, are locally used to constrain the prior Monte Carlo samples with well data. In contrast, the corresponding velocity of samples computed from a rock physics model is utilized to estimate the seismic data likelihood. Note that well data likelihood is performed locally at sparse well locations, while the seismic data likelihood is spatially distributed.   

    Consequently, the likelihood terms must be defined to retrieve the conditional realizations. Since these are high-dimensional nonlinear functions mapping inputs $\mathbf{b}$ to outputs $\mathbf{h}$, analytical likelihoods are impractical (except under unrealistic multivariate Gauss-linear assumptions). Therefore, both data conditioning is performed using approximate Bayesian computation (ABC). ABC technique is a family of statistical likelihood-free frameworks employed to perform Bayesian inference in complex problems \cite[]{Beaumont2010, Blum2010}. A detailed inspection of the theory behind practical ABC algorithms is reviewed in \cite{Brooks2011}. Moreover, \cite{Pradhan2020_SBEL} utilize ABC when inferring reservoir properties from seismic data. For our work, both likelihood computation for BGBM comes from ABC application. Summary statistics from composite likelihood methods \cite[]{Varin2011} is used to condition the models to the well data, while rejection sampling \cite[]{ABC_Jiang2017} is applied for seismic data calibration.

    The well data is measured in the same physical units as the simulated fields from basin modeling. Therefore, a weighted mismatch evaluation in multivariate form through marginal likelihoods is enough to estimate a metric for summary statistics. On the other hand, a similar assessment for the seismic data likelihood over all Monte Carlo samples requires enormous computational resources. One must forward simulate seismic data with wave propagation from the proposed velocity models of prior samples to apply the mismatch evaluation. Note that this practice is akin to the computationally expensive full waveform inversion (FWI) procedure to estimate velocity directly from seismic data \cite[]{Virieux2009}. Hence, a mismatch evaluation for the entire seismic data is computationally impractical for our purpose of analyzing the seismic data likelihood for thousands of Monte Carlo basin model samples.

    \cite{Pradhan2020_ABC} overcome this challenge by introducing the analysis of each Monte Carlo sample through seismic features extracted from the results of seismic kinematic models. Each Monte Carlo sample is reduced to a metric related to the flatness of reflection events after performing depth-migrations with the velocity from each realization. The authors use this metric as a proxy to evaluate the seismic data likelihood. However, the computation of depth-migration for thousands of Monte Carlo samples remains expensive and requires handling unmigrated pre-stack seismic data for each sample. Therefore, we propose two novel proxies for the seismic data likelihood evaluation using other kinematic models to reduce computational time. Furthermore, these proposed methods are even more practical as they only require the time-migration velocity spectra and the post-stack seismic image. All three proxies are detailed in the following sub-section.  
    
    \subsection{Kinematics criteria proxies used to analyze seismic data likelihood}

    The spatial likelihood term in BGBM is not computed with the seismic data per se because of the computational challenge in simulating the seismic data for each Monte Carlo sample. Therefore, kinematic models are used to analyze the seismic data likelihood of the realizations. These kinematic models use the acquired seismic data along with the velocity of a Monte Carlo sample computed through rock physics considerations. Then, kinematic properties are evaluated for each sample and reduced into a metric. Finally, this metric is used as a proxy for the rejection sampling strategy of ABC. 

    Therefore, the Monte Carlo samples which agree with seismic data are selected by appraising seismic kinematic features. The three cases discussed in this work are different proxies analyzing distinct seismic features. The forward kinematic modeling to reduce the velocity of a sample along with the seismic data into a metric is represented by $g_{KM}^{\left(i\right)}\left(.\right);\ \ i=$ benchmark, semblance, horizons. Figure \ref{fig:Figs/Figure2} illustrates the modeling procedure for each of the three proxies utilized in this work.

    \plot{Figs/Figure2}{width=\columnwidth}{\captionTwo}

    One must recall that seismic velocity estimation is a non-unique operation. For this procedure, generally, an optimization technique is used to retrieve a velocity model, which is a model from the $\mathbf{h}_\mathbf{v}\mid\mathbf{d}_{s,\mathbf{obs}}$ space. However, this model may not necessarily belong to the geologically consistent $\mathbf{h}_\mathbf{v},\ \mathbf{b}\mid\mathbf{d}_{s,\mathbf{obs}}$ space. In other words, the seismic velocity estimation does not assure a velocity model that is geologically consistent (and also in agreement with physical processes). On the other hand, the velocity generated from the seismic velocity estimation has desirable kinematic properties, as it flattens the gathers. Hence, the kinematic property of the legacy velocity for BGBM is used as a reference value for the chosen proxy when analyzing the seismic data likelihood of Monte Carlo samples. Note that we refer to the legacy velocity as the estimated seismic velocity used to define the original seismic image in which the inputs (e.g., horizons) for the basin modeling were outlined.

    The benchmark proxy ($\varepsilon_{benchmark}$), which is computationally expensive, is the one specified by \cite{Pradhan2020_ABC}. The metric generated by this kinematic model is considered a benchmark in this paper because it uses robust procedures and standard methods of model velocity analysis (MVA; \citeauthor{Biondi2006}, \citeyear{Biondi2006}) in each of its steps. First, the unmigrated acquired seismic data and the velocity associated with each Monte Carlo sample are used to perform a seismic depth-migration. Then, the flatness of the reflection events is evaluated in the angle domain common image gathers (ADCIGs; \citeauthor{Sava2003}, \citeyear{Sava2003}). This last procedure is compressed in the $\rho$ parameter extracted from the residual moveout (RMO) analysis  \cite[]{AlYahya1989}. Finally, all $\rho$ parameters of ADCIGs of the depth-migrated seismic traces are reduced to an averaged global flatness misfit represented by $\varepsilon_{benchmark}$.

    This global misfit value is the benchmark metric for the seismic data likelihood of realizations. The rejection sampling procedure using ABC permits defining a threshold value for this proxy to evaluate whether the sample belongs to the posterior (conditional) distribution. Therefore, the samples with acceptable seismic data likelihood have a lower proxy value than that computed using the legacy velocity. Figure \ref{fig:Figs/Figure3} illustrates these proxy values of samples for the real case application discussed in the next section. One rejected and one accepted sample is also selected to demonstrate the ADCIGs behavior in these realizations.
	    
    \plot{Figs/Figure3}{width=\columnwidth}{\captionThree}

    The computation of this benchmark proxy has two challenges: performing seismic depth-migration for several Monte Carlo samples and using unmigrated seismic data. Thus, two novel proxies are proposed utilizing different kinematic features. They are denoted as semblance proxy and horizons proxy. These new metrics were designed to overcome the use of expensive seismic depth-migrations and, hence, make the BGBM workflow computationally practical.
    
    Aspects of semblance proxy resemble the computation of structural uncertainty by time-migrated images proposed by \cite{Fomel2014}. Thus, in order to reduce the computational time by not performing several depth-migrations, this proxy utilizes MVA tools adequate for time-migration processes. In this case, the seismic data likelihood of a sample is specified by retrieving the semblance values obtained from the velocity of the realization. 
    
    The computation of semblance proxy ($\varepsilon_{semb}$) starts by rendering the semblance velocity scan from the unmigrated seismic data. The hypercube of this property is based on the normal moveout (NMO) procedure applied over a range of velocity values \cite[]{Yilmaz2001}. Therefore, each velocity associated with a Monte Carlo sample is transformed into root-mean-square (RMS) velocity in time domain to retrieve the semblance values for the realizations. This form of velocity represents a hypersurface crossing the semblance scan hypercube. Note that the approximation of RMS velocity to NMO velocity is explored. In order to generate a focused time-migrated seismic image, the velocity hypersurface should pass through high semblance values. 
    
    Consequently, the proxy for the semblance approach utilizes the semblance values extracted from the surface generated for each realization. Then, these values are reduced to an average value along all the spatial and time domains. Finally, we compute the inverse of this averaged value to serve as a metric for the seismic data likelihood of the realizations. The inverse value is taken for consistency with the other proxies, which use lower metric values as high likelihoods. 

    When utilizing this semblance proxy, the underlying assumption is that the area of interest presents only mild lateral variations in the velocity. Furthermore, the threshold value is retrieved from the legacy velocity to identify posterior samples in the rejection-sampling strategy. Figure \ref{fig:Figs/Figure4} demonstrates the values of realizations using the semblance proxy for the real case application discussed in the next section. Besides, one rejected and one accepted sample is also selected to illustrate the behavior of their RMS velocity curves in a semblance scan panel. 
        
    \plot{Figs/Figure4}{width=\columnwidth}{\captionFour}
    
    Note that the semblance proxy solely addresses the first challenge (i.e., computation) presented in the benchmark proxy. When computing this new proposed proxy, one still needs to handle unmigrated seismic data, which may not be commonly accessible to geomodelers who typically have only access to the post-stack seismic image information. Therefore, the horizons proxy is developed to tackle both challenges.
    
    The proposed horizons proxy is based on the fact that seismic images enable stratigraphic and structural interpretation away from well data. Furthermore, the structural inputs (horizons) for the basin modeling are typically retrieved from a focused seismic image. Consequently, the depth-positioning of present-day mapped interface structures - seismic horizons - embeds kinematic features of the seismic data. Therefore, the depth-positioning analysis of converted horizons corresponding to each Monte Carlo sample can be used to define a proxy for the seismic data likelihood of realizations.
    
    The depth-positioning analysis starts by using the legacy seismic velocity to convert the interpreted horizons from depth to time. Then, the computed velocity from rock physics modeling of Monte Carlo realizations is used to reconvert the seismic horizons from time to depth. These time-depth conversion procedures as a kinematic model are computationally fast and do not demand manipulation of unmigrated seismic data. Note that ray tracing \cite[]{Larner1981} could be selected as a conversion method. However, for simplicity, we have selected vertical conversion in our investigations. After all conversions of horizons to depth, the depth-positioning related to a Monte Carlo basin modeling sample is compared with the original depth of the seismic horizons using a Euclidean distance ($D_{hzs}$) in a reduced dimension space.
    
    Dimension reduction is performed using the classical multidimensional scaling (MDS) technique \cite[]{Wang2012}. The use of the MDS procedure in problems related to forecasting sub-surface properties for the hydrocarbon industry is found in \cite{Park2013}; and \cite{Scheidt2015}. Classical MDS and principal component analysis (PCA) are closely related as the first procedure assumes Euclidean metric. In this work, we deal with a data vector with elements representing the depth-positioning points of all the converted horizons. Each dimension in the classical MDS represents the projection of a data vector of a Monte Carlo sample onto one principal component from the matrix of the prior ensemble of converted horizons.
    
    Mathematically, each depth for prior sample $i$, horizon $j$, and lateral location $k$ is denoted by a scalar ${z_{j,k}}^{\left(i\right)}$. The indices run as follows: $i=1,\cdots,N_{prior}$; $j=1,\cdots,N_{hzs}$; and $k=1,\cdots,N_x$, where $N_{prior}$ stands for the number of prior samples, $N_{hzs}$ is the number of seismic horizons, and $N_x$ designates the number of lateral grid cells. Then the multivariate data vector for a single prior sample i is a row vector with dimensions $1\times\left(N_{hzs}\times N_x\right)$, as stated below:

    \begin{equation}
    \label{eq:vector_representation}
        \mathbf{z}^{(i)} = \left[ z_{1,1}^{(i)},\cdots,z_{1,N_{x}}^{(i)},\cdots,z_{N_{hz},N_{x}}^{(i)} \right]
    \end{equation}
    
    
    All the $N_{prior}$ $\mathbf{z}^{\left(i\right)}$ row vectors are stacked to form the prior ensemble data matrix, $Z$ with dimensions $N_{prior}\times\left(N_{hzs}\times N_x\right)$. Principal component analysis is done on this data matrix to give $P=ZV$, where the columns of $V$ are the eigenvectors (principal components - PCs) and $P$ elements are the scores. To perform classical MDS, one takes the first $m$ PC scores $\left(p_1,\cdots,p_m\right)$ for each prior sample $i$ and computes all pairwise Euclidean distances. One detail is that the depth points of each data vector are not weighted uniformly but exponentially, inversely proportional to the original depth. This aspect is described later in the discussion section. Thus, the PCA gives $P=\widetilde{Z}V=ZWV$, where the original vector $\mathbf{z}^{\left(i\right)}$ is transformed to ${\widetilde{\mathbf{z}}}^{\left(i\right)}$ by weight multiplication on each vector element. The weights are represented as a diagonal matrix $W$ with dimensions $\left(N_{hzs}\times N_x\right)\times\left(N_{hzs}\times N_x\right)$.
    
    For this work, the number $m$ of reduced dimensions considered is such as to span 99\% of the prior variability in the depth-positioning of the resultant horizons. Then, the MDS representation of each sample through the converted horizons feature requires only m dimensions (reduced from $N_{hzs}\times N_x$) with the first m principal component scores. Specifically, $m=6$ for the data of the real case application, while $\left(N_{hzs}\times N_x\right)= 10\times600=6000$. Finally, the Euclidean distance of the converted horizons of a sample data vector from the original depth-positioning of the horizons is computed in this reduced-dimensional space. This distance in the classical MDS space represents the horizons proxy ($\varepsilon_{hzs}= D_{hzs}$) for seismic data likelihood.
    
    Note that this horizons proxy bears two underlying assumptions: the legacy velocity model yields a focused seismic image in the area of interest, and the horizon interpretation realistically reflects the stratigraphic geological layering. The last assumption indicates that the seismic data information is contained in the seismic horizons. The resulting structure of mapped stratigraphic events through amplitude reflectivity tends to follow the morphology of vertically adjacent seismic reflection patterns of the seismic image. 
    
    According to \cite{Pon2005}, slight variations in velocity can cause small perturbations in the depth of seismic migrated data. Therefore, when a velocity model with changes from the legacy produces a clear seismic image and is used to convert time to depth, one may observe only minor perturbations in the depth of stratigraphic horizons. Thus, the statistical distance $\varepsilon_{hzs}$ is justified as a metric to analyze the seismic data likelihood with some acceptable degree of variation in the distance value. Therefore, a distance threshold is used in the ABC rejection sampling methodology to accept the basin models with horizons that are “close” to the original depth-positioning of horizons in the reduced dimension space. The definition of the distance threshold for this proxy is heuristic. This reference distance is computed by creating hypothetical depth values from the original depth of seismic horizons by adding a percentage perturbation to the original depths. The selected percentage represents the maximum expected error in the interpreted depth-positioning value. Thus, the seismic data likelihood evaluation accepts any sample with a distance below this threshold. In this work, the heuristic margin for the misfit of depth values for the horizons is 2\%.
    
    Figure \ref{fig:Figs/Figure5} exhibits, in the first two dimensions of the MDS space, the horizons proxy as colors for the realizations of the real case application discussed in the next section. Besides, one rejected and one accepted sample is also selected to illustrate the behavior of the resultant depth-positioning of the horizons compared with their original depth-positioning. The axes of the cross-plot are the first two principal dimensions of the MDS, in which the $\alpha_i$ value represents the normalized variance of priors covered by the $i^{th}$ dimension.
    
    \plot{Figs/Figure5}{width=0.93\columnwidth}{\captionFive} 
    
    In summation, the benchmark proxy is the most robust. However, it is the most computationally time-consuming. The proposed semblance proxy is computationally fast and robust if the area of interest meets the assumption of no strong lateral variations in velocity. Nevertheless, it still needs the use of unmigrated seismic data. Finally, assuming a focused seismic image when designing the structural horizon information as basin modeling inputs, the horizons proxy is proposed. Both challenges presented for the benchmark case are tackled using the latter proxy as a metric to select posterior earth models conditioned to seismic data. The analysis for the horizons proxy is computationally fast and does not require using unmigrated seismic data. The next section compares results for pore pressure prediction and its uncertainty quantification using the three described proxies in the BGBM workflow. 


\section{Real Case Application}

The BGBM workflow is applied to real data to forecast pore pressure along with uncertainty quantification. In order to examine and compare the pore pressure prediction results from the three proxies for the seismic data likelihood evaluation of realizations, the same data and geological assumptions studied by \cite{Pradhan2020_ABC} were considered in this work. The investigated area is located off the coast of Louisiana in the north-central Gulf of Mexico (GoM) basin. The analyzed 2D section crosses the blocks of Ship Shoal and South Timbalier. The geological evolution of the GoM basin is well described in \cite{Galloway2008}.

The area of interest presents a high sedimentation rate for the Quaternary and Neogene geological periods. Disequilibrium compaction is the primary mechanism for overpressure in areas where sediments are swiftly deposited \cite{AAPGmemoir70_ch1}. At the location of our studies, rocks from the Miocene Epoch may be found up to 10 kilometers in depth. Hence, pore pressure prediction becomes essential for this area in order to avoid harm to the environment and ensure human safety when designing drilling projects.

\plot{Figs/Figure6}{width=0.9\columnwidth}{\captionSix}

Figure \ref{fig:Figs/Figure6} illustrates the 2D section employed in our workflow appraisal. The white horizon represents the top of the salt interpretation, and this surface was considered to remain unchanged during the sedimentation of the layers above it. The analyzed section crosses well-1 (magenta line), where porosity data obtained from bulk density log, mudweight measurements, and bottom-hole temperature values were considered for basin modeling calibration. Moreover, logs from this well were used to calibrate deterministic parameters for the rock physics velocity transformation, such as end-member mineral moduli. Furthermore, a projection of a blind well is illustrated in the figure sections by a gold line. The mudweight data from this blind well, which is located 1.6 km away from the 2D section, is used to validate the workflow results for the posterior pore pressure forecast. The blind well projection on the studied section is located more than 15 kilometers away from well-1.

\subsection{Establishing the prior uncertainty}

    For basin modeling simulations, the prior uncertainty is established in the geological inputs, $\mathbf{b}$. Theoretically, all possible geological inputs could be considered uncertain. However, if every possible input is considered uncertain, the number of dimensions in the vectorial space of the problem would be cumbersome for Monte Carlo sampling \cite[]{QUSS_book}. Fortunately, previous geological expertise in the area of interest has facilitated the selection of parameters to be considered either uncertain or deterministic.
    
    The objective, in this case, is to predict the pore pressure within the shaley-sand layers of the examined section. Therefore, facies distribution and facies parameters that are neither related to porosity nor permeability are considered deterministic. On the other hand, facies parameters that control the mechanical compaction \cite[]{Athy1930} and porosity-permeability relations \cite[]{Ungerer1990} must be considered uncertain due to their control of fluid flow during the geological time. Hence, prior probability distributions are assigned to them. 

    This study considered one homogeneous facies for each of the ten interpreted stratigraphic layers. Appendix A reports the values established for the prior distributions of these parameters. In any Monte Carlo study, a preliminary step to make sure that the priors are appropriate is to test that the data simulated from the priors are not statistically inconsistent with the observed data - a step sometimes called as prior falsification. This statistical test to make sure that the priors are not falsified was performed for the selected priors. Details are in Appendix A. In addition, the reader is encouraged to consult \cite{Pradhan2020_ABC} for more details about how the prior distributions were established.

    Clay minerals in GoM are expected to suffer from diagenesis (smectite to illite transformation). Diagenetic processes also affect pore pressure results. Hence, parameters controlling diagenesis must be treated as uncertain. The approach followed in this work is described by \cite{Dutta1987} and associates the chemical diagenesis of smectite-illite mineral transition in clay sediments with the thermal history of the basin. A first-order Arrhenius rate theory simulates the chemical reaction controlling the smectite-illite transition in basin modeling. Therefore, the boundary condition value of heat flow during the sedimentation of the geological layers was also considered as an uncertain parameter. The heat flow value was assumed as spatially constant throughout the geo-history in the basin modeling.           
    
    In summation, the uncertain prior model space, $\mathbf{b}$, for the basin in this area of interest has 51 parameters with established broad distributions: two parameters related to mechanical compaction law (Athy's coefficient in depth units and initial porosity of sedimentation) and three parameters for porosity-permeability relationship (permeability scale factor, specific surface, and anisotropic factor) for each of the ten layers, which is one facies per layer. Additionally, the basal heat flow value as a boundary condition is uncertain. Any other input parameter necessary to simulate basin models was defined as deterministic and assigned a value by standard techniques in basin modeling workflows.

    Monte Carlo sampling is performed using these prior distributions to generate 2,500 basin modeling outputs, $\mathbf{h}_{-\mathbf{v}}$ from simulations. As a result, prior realizations of pore pressure, vertical stress, porosity, temperature, and clay mineral distributions become available for further data assimilation. Note that we consider the clay volume in each layer as constant, which was computed from the well-1 gamma-ray log (see Appendix A). The basin modeling was performed using the PetroMod commercial software. To proceed with the workflow, rock physics modeling is used to relate the basin modeling outputs to velocity models, $\mathbf{h}_\mathbf{v}$.

\subsection{Rock physics modeling}

    Rock physics is the link between rock properties and geophysical data. Therefore, rock physics modeling is used in the workflow to generate velocity from basin modeling outputs. Velocity models are the most sensitive parameter in seismic data evaluation, such as migration procedures to yield seismic images. Hence, geologically-consistent velocities from rock physics modeling using outputs from basin models are the key to assimilating seismic data in the BGBM workflow.
     
    \plot{Figs/Figure7}{width=\columnwidth}{\captionSeven}
    
    \cite{Brevik2011} claims that rock models should be universal. Hence, a set of basin modeling properties at a sub-surface point can generate the corresponding velocity value along with rock physics modeling. Furthermore, porosity-velocity relationships \cite[]{book_RPH_2009} demand information that can be generated by basin models and inferred from well-logs. Therefore, simulated pore pressure ($P_p$), stress ($S$) history, temperature, fluid saturation, and mineral distribution of the rocks are associated through rock physics models to compute the velocity field of prior Monte Carlo samples.

    One may generate sub-surface fields ($\mathbf{h}_{-\mathbf{v}}$) from basin modeling by Monte Carlo sampling of the $\mathbf{b}$ vector. Then, these simulated properties with an appropriate rock physics model under some assumptions are used to generate a velocity model consistent with geological constraints. Finally, a complete set of prior field hypothesis $\mathbf{h}_\mathbf{g}=\left(\mathbf{h}_{-\mathbf{v}},\mathbf{h}_\mathbf{v},\mathbf{b}\right)$ is obtained for each sample. Figure \ref{fig:Figs/Figure7} illustrates this procedure. For the particular case of this study, the shaly-sand model \cite[]{Avseth2005} is adopted. 
    	
\subsection{Well data assimilation}

    Ideally, data assimilation should be done synchronously for all types of data. However, solving a joint Bayesian inference problem with different types of generative models and diverse data to condition is computationally challenging and necessitates using reasonable approximations and smart sampling strategies. Therefore, the importance sampling (IS) technique \cite[]{Book_MC_strategies} is employed in this study to avoid several unnecessary time-consuming depth migrations when evaluating the seismic data likelihood with benchmark proxy. Furthermore, the use of IS procedure is standard when utilizing ABC rejection sampling \cite[]{Handbook_of_ABC}. Moreover, this strategy was also adopted by \cite{Pradhan2020_ABC}. 

    The conditional independence assumption between the well and seismic data allows the joint posterior distribution to be written as a product of posterior distributions. Consequently, the data calibration may be done sequentially rather than jointly, as the posterior samples must have both acceptable well data and seismic data likelihoods. Hence, to reduce the seismic calibration calculations, IS strategy is used after the well data conditioning to draw pre-posterior samples that are already close to well data. 
    
    The well-1 data used for calibration in the real case experiment consists of porosity, mudweight, and bottom-hole temperature data at specific depths. The ABC procedure to condition models with well data selects a small number of samples through summary statistics analysis. For our case, an auxiliary likelihood \cite[]{Handbook_of_ABC} is used to compute the mismatch of the real well-1 data with the simulated models. The ABC method enables generating posterior samples in cases where the likelihood cannot be analytically expressed or is intractable to compute.

    Even though one may write an explicit well-1 data likelihood under Gaussian assumptions, the Gaussian function is unsuitable for expressing the marginal likelihood of pore pressure. Moreover, the correlation matrix would also be required. Therefore, for simplicity and to address the pore pressure likelihood design, we have defined marginal likelihoods for each data point of the well with shapes accordingly to the physical meaning of the measurement. Next, the auxiliary likelihood ($f_{aux}(\mathbf{d}_{\mathbf{w}, \mathbf{obs}}\ \mid\mathbf{h}_{-\mathbf{v}},\mathbf{b})$) that acts as a summary statistic is assembled by multiplying the marginal likelihoods. This composite likelihood for ABC used in our example assumes orthogonality between the measured values, and it is known as the independence likelihood \cite[]{Chandler2007}. Finally, a logarithm mismatch between the observed data and the simulated data values from summary statics is computed as metric ($\varepsilon_\mathrm{w}$).

    For the marginal likelihoods, the observed data vector $(\mathbf{d}_{\mathbf{w},\mathbf{obs}} =[\mathbf{d}_{\mathbf{w},\mathbf{obs}}^{(\phi)}, \mathbf{d}_{\mathbf{w} ,\mathbf{obs}}^{(P_p)}, \mathbf{d}_{\mathbf{w}, \mathbf{obs}}^{(T)}]^t)$ is decomposed into small vectors by physical properties. Then, Gaussian distributions centered in the individual measured value to porosity data ($\mathbf{d}_{\mathbf{w},\mathbf{obs}}^{(\phi)}$) and corrected bottom-hole temperature data ($\mathbf{d}_{\mathbf{w},\mathbf{obs}}^{(T)}$) with a standard deviation of $\sigma_\phi=0.01$ and $\sigma_T=2^oC$ were respectively assigned as marginal distributions. The deviations express uncertainty in the data acquisition and temperature correction procedure. Moreover, a truncated triangular distribution with a maximum probability assigned to the recorded mudweight values ($\mathbf{d}_{\mathbf{w},\mathbf{obs}}^{(P_p)}$) was considered for the pore pressure marginal likelihood. Unlike the likelihoods for porosity and temperature which are symmetric with respect to the measured value, the pore pressure likelihood is asymmetric with respect to the measured mudweight since the mudweight is an upper-bound estimate of the actual pore pressure. The mudweight is usually kept higher than the actual pressure of the formation. Therefore, the recorded mudweight values were taken as upper limits on the pore pressure likelihood. The marginal likelihood region greater than the mudweight value is assigned a zero value. Moreover, as hydrostatic pressure ($P_{hyd}$) can be thought of as the lower limit on the pore pressure at any given depth, this pressure is assigned as the minimum probability in the truncated triangular marginal likelihood for pore pressure. 

    The letter $N$ with each of the physical properties $l$ ($l\in\{\phi,P_p,T\}$) subscribed denotes the total number of measured points for the next equations. The auxiliary likelihood for each property is the multiplication of each marginal likelihood related to a measurement, as presented below:  

    \begin{equation}
    f_{aux}\left(\mathbf{d}_{\mathbf{w},\mathbf{obs}}^{(l)}\mid\mathbf{h}_{-\mathbf{v}},\mathbf{b}\right)=\prod_{i}^{N_l}{f\left(\left[\mathbf{d}_{\mathbf{w},\mathbf{obs}}^{(l)}\right]_\mathbf{i}\mid\mathbf{h}_{-\mathbf{v}},\mathbf{b}\right)};\ \ \ l\in\{\phi,P_p,T\}
    \end{equation} 

    Equations \ref{eq:marg_phi}, \ref{eq:marg_pp}, and \ref{eq:marg_T} respectively represent the auxiliary likelihoods of porosity, pore pressure, and temperature:
    \begin{equation}
    \label{eq:marg_phi}
        f_{aux}\left(\mathbf{d}_{\mathbf{w},\mathbf{obs}}^{(\phi)}\mid\mathbf{h}_{-\mathbf{v}},\mathbf{b}\right)=
        \prod_{i}^{N_\phi}{\left(2\pi\sigma_\phi^2\right)^{-1/2} \exp{\left[-0.5\sigma_\phi^{-2}\left(\left[\mathbf{d}_\mathbf{w}^{(\phi)}\right]_\mathbf{i}-\left[\mathbf{d}_{\mathbf{w},\mathbf{obs}}^{(\phi)}\right]_\mathbf{i}\right)^2\right]}}
    \end{equation} 

    \begin{eqnarray}
    \label{eq:marg_pp}
        f_{aux}\left(\mathbf{d}_{\mathbf{w}, \mathbf{obs}}^{(P_p)}\mid\mathbf{h}_{\mathbf{-v}},\mathbf{b}\right) = \begin{cases}
        \prod_i^{N_{P_p}} \frac{2([\mathbf{d}^{(P_p)}_{\mathbf{w}}]_i-[\mathbf{P}_{hyd}]_i)}{([\mathbf{d}^{(P_p)}_{\mathbf{w,obs}}]_i-[\mathbf{P}_{hyd}]_i)^2} \;&\mbox{if~} [\mathbf{d}^{(P_p)}_{\mathbf{w}}]_i \in [[\mathbf{P}_{hyd}]_i,[\mathbf{d}^{(P_p)}_{\mathbf{w,obs}}]_i] \;\forall i \\ \\
        \;\;\;\;\;\;\;\;\;\;\;\;\;\;\;\;\;\mathbf{0}\;&\mbox{else} \end{cases}
    \end{eqnarray}
    
    \begin{equation}
    \label{eq:marg_T}
        f_{aux}\left(\mathbf{d}_{\mathbf{w},\mathbf{obs}}^{(T)}\mid\mathbf{h}_{-\mathbf{v}},\mathbf{b}\right)=
        \prod_{i}^{N_T}{\left(2\pi\sigma_T^2\right)^{-1/2} \exp{\left[-0.5\sigma_T^{-2}\left(\left[\mathbf{d}_\mathbf{w}^{(T)}\right]_\mathbf{i}-\left[\mathbf{d}_{\mathbf{w},\mathbf{obs}}^{(T)}\right]_\mathbf{i}\right)^2\right]}}
    \end{equation}  
        
    Then, the auxiliary likelihood used as summary statistics for ABC (${S(\mathbf{d}}_\mathbf{w})$) is stated below in equation \ref{eq:summary}. Note that this statistical value is composed as the multiplication of each auxiliary likelihood:

    \begin{equation}
    \label{eq:summary}
        {S(\mathbf{d}}_\mathbf{w})=f_{aux}(\mathbf{d}_{\mathbf{w},\mathbf{obs}}\mid\mathbf{h}_{-\mathbf{v}},\mathbf{b})=\prod_{l\in\{\phi,P_p,T\}}{f_{aux}\left(\mathbf{d}_{\mathbf{w},\mathbf{obs}}^{(l)}\mid\mathbf{h}_{-\mathbf{v}},\mathbf{b}\right)}
    \end{equation}

    Finally, the function (used as metric for well data) measuring the logarithm mismatch between simulated and observed points ($D_{\mathbf{d}_{\mathbf{w},\mathbf{obs}}}({S(\mathbf{d}}_\mathbf{w}))\ \equiv\varepsilon_\mathrm{w}$) is established below as equation \ref{eq:wmetric}: 

    \begin{equation}
    \label{eq:wmetric}
        \varepsilon_\mathrm{w} = D_{\mathbf{d}_{\mathbf{w},\mathbf{obs}}}({S(\mathbf{d}}_\mathbf{w}))=\log\frac{{S(\mathbf{d}}_{\mathbf{w},\mathbf{obs}})}{{S(\mathbf{d}}_\mathbf{w})}=\log{\ S(\mathbf{d}}_{\mathbf{w},\mathbf{obs}})-\log\ {S(\mathbf{d}}_\mathbf{w})\ 
    \end{equation}	

    After computing the misfit from the composed likelihood for 2,500 prior samples, 30 models with the lowest mismatch (low $\varepsilon_\mathrm{w}$) against well-1 data were selected. These selected models represent samples conditioned only to well-1 data (no seismic data conditioning yet). The practical ABC approach of retaining few realizations of the priors as models approximately conditioned to data through summary statistics analysis is described in \cite{Beaumont2002}.

    Part (a) of Figure \ref{fig:Figs/Figure8} depicts vertical outputs of porosity, pore pressure, and temperature from these 30 models (black curves) and prior models (gray curves). The black dots with white rims represent the measured data on well-1. Kernel Density Estimator (KDE) is applied to the uncertain basin modeling input parameters $\mathbf{b}$ of the 30 selected models to define pre-posterior distributions for IS. The marginal distributions of each uncertain parameter constitute this newly proposed pre-posterior distribution. As an illustration, the distributions of uncertain basin modeling input parameters ($A_c$: Athy’s coefficient, $\phi_0$: Initial porosity at deposition, $S$: Specific surface area, $F$: Permeability scaling factor, and $a_k$: Permeability anisotropy factor) of the seventh layer and the heat flow value are shown in part (b) of the figure. Again, the gray curves are related to the prior distribution, whereas the black ones represent the IS pre-posterior distributions for resampling.
	
    \plot{Figs/Figure8}{width=0.8\columnwidth}{\captionEight}

    The IS strategy, in this case, supports the claim that samples derived from its distribution have higher chances of matching the well-1 data. Consequently, this allows generating a more significant number of models approximately conditioned to well-1 data after IS resampling. The well data assimilation for the new IS samples is similar to that of model selection from prior sampling as the same summary statistics computation is used. 

    Therefore, 10,000 new basin model realizations were performed sampling from the pre-posterior distributions to retain 1,000 models from these IS samples with the lowest misfits against well-1 data. These 1,000 samples are considered as calibrated to well-1 data and they undergo a seismic data likelihood evaluation through ABC rejection sampling. The rejection sampling for seismic data calibration is based on one of the kinematics metrics described in the methodology section. Finally, these last accepted samples are conditioned to well and seismic data and belong to the posterior distribution.

    Figure \ref{fig:Figs/Figure9} summarizes the IS strategy implementation in the real case example. The MDS reduced space used to compute horizons proxy is utilized to aid in visualizing the evolution of the behavior of samples. Part (a) of the figure illustrates the 2,500 samples from the prior distribution in this MDS space as colored dots. The color represents a normalized Euclidean distance in this space between the sample and the original depth-positioning point, which is represented by red diamonds in the figure. In part (b), only the selected 30 models with the highest well data likelihood are highlighted. These few samples are used to estimate the pre-posterior distributions used for IS. Finally, part (c) shows the 1,000 samples with the highest well data likelihood among the 10,000 resamples using the pre-posterior distributions. 
	
    \plot{Figs/Figure9}{width=\columnwidth}{\captionNine}

    Note that samples after IS scheme present in part (c) of the figure depict a smaller cloud size than that formed by prior distribution samples shown in part (a). However, several points far away from the original depth-positioning point remain after IS. Consequently, one expects a further reduction of this cloud after seismic data assimilation. The following sub-section presents these posterior BGBM samples after seismic data likelihood evaluation.
\subsection{Posterior samples from kinematics criteria} 

    The well data likelihood computation was approximated using the data misfit with the appropriate function (Gaussian or truncated triangular) for each measurement. The samples conditioned to well data are selected based on the summary statistic of realizations (misfit to well-1 data using auxiliary likelihood). However, the application of mismatch for seismic data likelihood is cumbersome even for a single realization. Hence, proxies were employed to select models calibrated with seismic data. All three proxies explained in the methodology section were implemented to evaluate the seismic data likelihood in the remaining 1,000 instances generated from IS strategy and conditioned to well data.

    The thresholds used for the proxies to employ ABC rejection sampling in order to calibrate models with seismic data were either selected from the proxy value generated by the legacy velocity (benchmark and semblance proxies) or by heuristic means (horizons proxy). The estimation of these threshold values is explained in the methodology section. Moreover, illustrations of the proxy values and the thresholds can also be encountered in the mentioned section.

    The MDS space is one convenient way to illustrate and compare the posterior selection using different means for seismic data calibration. The accepted samples under the three described criteria are shown in Figure \ref{fig:Figs/Figure10}. Each plot highlights points by coloring the posterior samples with the horizons proxy value according to each seismic data calibration case.

    \plot{Figs/Figure10}{width=\columnwidth}{\captionTen}

    The selected posterior realizations have acceptable well data and seismic data likelihood, whereas the latter is according to the used proxy. The retrieved insight from Figure \ref{fig:Figs/Figure10} is that the posterior samples cluster near the original depth-positioning of the seismic horizons (red diamond) for all three proxies. Nevertheless, the semblance proxy admits more posterior points than the others. We explore this topic in the discussion section. In addition, the pore pressure fields associated with the posterior samples are used to predict and quantify the uncertainty of this property at locations away from well-1.
    
\subsection{Pore pressure prediction at blind well}

    This sub-section exhibits the pore pressure prediction results utilizing the different proxies for the seismic data misfit evaluation. For that, posterior realizations of the predicted pore pressure are retrieved on a vertical projection of the blind well in the studied section. Then, the posterior probability distributions were computed by KDE at a depth where mudweight measurement is available in the blind well. Only the mudweight at shallow depth is considered for workflow validation, as the deeper part of the blind well crosses a fault. Therefore, the deeper mudweight measurements and their projection are geologically different from the studied section.

    Figure \ref{fig:Figs/Figure11} depicts the vertical posterior realizations of pore pressure on the blind well projection for all proxies and the prior realizations. The right plot on the figure illustrates the marginal probability density functions at depth retrieved from three tiers along the BGBM workflow: the prior samples (gray), the posteriors with only well-1 data calibration (blue), and posterior realizations under different seismic data likelihood proxies (orange – benchmark, purple – semblance, and green – horizons).

    \plot{Figs/Figure11}{width=\columnwidth}{\captionEleven}

    The figure illustrates pore pressure prediction with uncertainty quantification before drilling (predrill stage). The depth-position analysis point is the depth of the mudweight measurement available from the blind well with similar geomorphology to the studied section. The mudweight value is represented by the red dashed line. Recall that the pore pressure value is expected to be close to and lower than the mudweight value. The peak of the posterior distributions has shifted from the prior distribution peak towards the mud-weight measurement, which validates the BGBM application for pore pressure prediction with uncertainty quantification. 

    Moreover, all the proxies show significant results, especially the benchmark and horizons proxies. The addition of seismic data calibration reduces uncertainty compared to the distribution in which only the offset well (well-1) data was assimilated. Therefore, incorporating exhaustive spatial seismic data in basin simulations benefits the forecast. Discussions related to how the proxies for seismic data likelihood affect the pore pressure prediction, particularly for this study, are reserved for the next section.


\section{Discussion}

This section explores issues related to the behavior of the proxies and encountered nuances when performing the experiment described in previous sections. Finally, an additional topic is addressed to summarize the obtained results by utilizing the three different proxies for the seismic data likelihood described in the methodology section.
    
\subsection{Why not use the velocity as a proxy?}

    BGBM aims to explore subsurface realizations with associated velocities not typically reached by seismic velocity estimators. Therefore, the velocity of a sample is likely to be different from the legacy velocity. The goal is not to match the legacy velocity but to explore additional velocity models consistent with the seismic data, as well as consistent with the geohistory and physical processes of a basin. Consequently, the velocity of a realization must be constrained with characteristic kinematic features as observed in the seismic data rather than just reproducing the same legacy velocity. 

    \plot{Figs/Figure12}{width=\columnwidth}{\captionTwelve}

    In the case of using statistical distance in the depth-positioning of converted seismic horizons, the velocities only need to yield similar reflection patterns observed on the original seismic image to be accepted as a posterior sample in the workflow. As the depth-positioning of the stratigraphic horizons tends to follow similar reflection patterns as seen in seismic images, then the double conversion of horizons with the legacy velocity and the velocity from samples must produce similar horizon depth-positioning to have acceptable seismic data likelihood. 
    
    The interpreted seismic horizons embed information about the seismic reflection patterns that the velocity field statistics would not carry. Moreover, if the statistical reduction of dimensionality were performed with the whole velocity field, the results would be prone to include noise of areas in which the modelers have no interest – for instance, velocity values at depths below the last stratigraphic horizon of interest. 

    Furthermore, the depth-positioning of horizons is represented by sparse spatial points, whereas the sub-surface velocity field covers all spatial coordinates in a grid. Hence, the statistical analysis using seismic horizons is more practical, stable, and computationally faster than the whole velocity field statistics.  

    Despite all these arguments, similar statistical analysis from horizons proxy computation was employed based on velocity fields to gather intuition about a possible proxy established by velocities. Figure \ref{fig:Figs/Figure12} illustrates cross-plots of the benchmark proxy value with the computed Euclidean distance of Monte Carlo samples in the reduced space of prior samples using different styles of velocity field as the reduced feature. The origin of distance calculations is related to the legacy velocity model, i.e., distance is zero at the legacy velocity model point in the reduced space. The styles of velocity field considered as features presented in the figure are: (a) interval velocity – with proxy labeled as $\varepsilon_{vp,int}$, (b) slowness – proxy as $\varepsilon_{vp,c}$, (c) RMS velocity in time domain – proxy as $\varepsilon_{vp,rms}$, and (d) averaged velocity along the depth axis – proxy as $\varepsilon_{vp,z-avg}$. The samples used for this evaluation are the same 1,000 realizations from IS after well data calibration shown in the previous section.
    
    On the one hand, the proxy considering the interval velocity for statistical distances seems severely uncorrelated with the benchmark proxy. On the other hand, the remaining proxies with different velocity styles demonstrate a vague linear trend with the benchmark proxy values. However, this correlation is lost in the area of posterior selection by the benchmark proxy – points below the dashed line (benchmark proxy threshold). Therefore, selecting a threshold for these velocity proxies to include the benchmark posterior samples would also retain several points with a high $\varepsilon_{benchmark}$ value. Consequently, using the statistical analysis on velocity fields to define proxies for the seismic data likelihood of samples is not recommended.
    
    Other insights are also retrieved from this experiment. Even though the velocities represented by the points below the dashed line in Figure \ref{fig:Figs/Figure12} have considerable distance from the legacy, they return focused seismic images after depth-migration. The velocity estimation from seismic data is a non-unique process. Therefore, these cross-plots illustrate that different velocities can honor kinematic features. In addition, one may note a remarkable similarity in the cross-plot point pattern that considers the averaged velocity along the depth axis and the slowness field. This resemblance indicates that the benchmark proxy and horizons proxy are also alike.

\subsection{Bias in semblance proxy}

    The GoM basin presents a dominant vertical gradient only with mild variations \cite[]{Fomel2014}, which justifies using time-migration tools along with velocity fields as a proxy for seismic data likelihood. However, most input parameters for basin modeling and the proxy threshold definition are based on the legacy velocity model obtained from a depth-migration. Therefore, time-migration kinematic property evaluation introduces bias in the posterior selection of samples. Recall that the semblance velocity spectra computed from the NMO procedure (based on a hyperbolic travel-time approximation) were used for semblance proxy. 

    \cite[~p.~S100]{Iversen2008} says: “historically speaking, ‘ideal’ or ‘complete’ time migration has been considered a process positioning the reflecting horizons ‘vertically above’ their corresponding reflectors in the depth domain.”. In order to investigate if the latter commentary holds in our real case application, Figure \ref{fig:Figs/Figure13} was elaborated.

    \plot{Figs/Figure13}{width=\columnwidth}{\captionThirteen}

    The figure illustrates converted horizons ensembles of posteriors (gray lines) in the spatial domain without applying any dimensional reduction. The ensembles of posteriors were generated by (a) benchmark and (b) semblance proxies. In addition, the original depth-positioning of horizons (black lines) is also depicted in the figure for reference. The applied round time-depth conversions in the seismic horizons for posterior samples follow the same procedure in creating the kinematic feature for horizons proxy.

    Analyzing Figure \ref{fig:Figs/Figure13}, one may note that the ensemble of posterior results for semblance proxy exhibits shallow depth-positioning bias with respect to the original depth-positioning. On the other hand, the benchmark proxy generates horizons with depth-positioning similar to the original ones.
    
    Furthermore, Figure \ref{fig:Figs/Figure14} strengthens the cited argument showing the legacy velocity transformed to time domain RMS velocity as the black dashed curves overlaying the semblance spectra at specific common midpoints (CMPs). It is observed in the figure that the black curves are slightly shifted to the right from the highest semblance values. Hence, the posterior sample selection for semblance proxy will be skewed to the ones with lower velocities. In addition, this skewness explains why more samples were accepted as posteriors for semblance proxy compared to the other proxies (see Figure \ref{fig:Figs/Figure10}).

    \plot{Figs/Figure14}{width=\columnwidth}{\captionFourteen}
    
    Besides, this bias in the semblance proxy posterior selection is reflected in the pore pressure results. The posterior selected models admit either more permeable and higher porous rocks with lower pore pressure or lower permeable and less porous rocks with higher pore pressure when compared to the benchmark result. (see Figure \ref{fig:Figs/Figure11}). These geological arguments are related to velocities with lower values. Thus, the pore pressure distribution uncertainty is more significant than the ones retrieved by our benchmark proxy and horizons proxy. Consequently, using kinematic properties based on time-migration procedures as a proxy for the seismic data likelihood for samples with geological velocities derived using parameters from depth-migrated images is not recommended.

\subsection{Weights in horizons proxy}

    When doing principal component analysis, the results depend on the units or scales of the different components of the multivariate data vector. Here, the velocity averaged along the depth axis follows a gradient as one deals with vertical spatial sections, and the different horizons have a large variation in depth. The deepest horizon has much larger depth values than the shallowest. Hence, the distance based on the depth-positioning feature should reflect a depth-related weight in its statistical principal component analysis and reduction. Therefore, instead of the regular standardization with uniform weights when realizing the dimensionality reduction \cite[]{QUSS_book}, weights proportional to depth must follow in the computation of horizons proxy values.

    \plot{Figs/Figure15}{width=\columnwidth}{\captionFifteen}
    
    Moreover, the correct positioning of shallow structures dictates the overall performance of seismic migration \cite[]{Jones2012}. Therefore, the depth-positioning analysis of seismic horizons must consider a weighting scheme in which the shallowest ones contribute more than the deeper horizons. Furthermore, seismic imaging quality decreases with depth. Hence, shallower horizons get more weight as horizon picking depends on the legacy seismic image. 

    In order to analyze these arguments, an experiment was conducted when computing the horizons proxy. Two weighting schemes on the converted depth-positioning of seismic horizons were performed during the dimension reduction procedure. The first (the method used for the horizons proxy in the real case application) considers an exponential weight decreasing with depth – the inverse value of the depth of the original depth-positioning as a weight for each horizon point. The second strategy employs a uniform weight for all points of horizons - the regular standardization when a uniform mean is expected. Equations for each diagonal element of the weight matrix for both types of weighing are shown below:

    \begin{equation}
    \label{eq:exp_weights}
        w_i^{\left(exponential\right)}=\frac{K}{z_i} \;;\;\; \sum_{i}^{N_{hzs} \times\ N_x}w_i = \frac{N_{hzs}N_x}{\sqrt{N_{priors}}}
    \end{equation}

    \begin{equation}
    \label{eq:reg_weights}
        w_i^{\left(regular\right)} = \frac{1}{\sqrt{N_{priors}}}
    \end{equation}

    Equation \ref{eq:exp_weights} represents the weights for the exponential case applied in the features to perform dimensional reduction, and Equation \ref{eq:reg_weights} is the case with constant weights (regular standardization). In these equations, $N_{hzs}$ represents the number of horizons, $N_x$ is the number of cells along the horizontal section direction, and $N_{priors}$ is the number of prior samples ($N_{priors} = 2,500$ in this case). Note that the number of points in the horizons depth-positioning feature for a single sample is $N_{hzs} \times N_x$. Moreover, $z_i$ represents the depth of the $i^{th}$ point of the original seismic horizons, and $K$ is a normalizing constant determined by the constraint on the sum of the weights in the exponential case.
    
    Part (a) of Figure \ref{fig:Figs/Figure15} illustrates the pore pressure results at the blind well measurement for the two weighting schemes. The same heuristic cut-off described in the methodology section for the horizons proxy was considered in both cases. Moreover, the posterior converted horizons ensembles are also shown to compare these scheme results in parts (b) and (c) of the figure.
    
    The resultant depth-positioning ensemble of posteriors using regular standardization illustrated in part (c) presents wider variations for the shallowest horizons when compared with exponential weights shown in part (b). Moreover, when using the uniform weighting scheme, the pore pressure results at the validation point exhibit an unexpected bimodal distribution. Therefore, one must consider higher weights for shallow horizons when computing and projecting onto the reduced space for the distance computation be a reliable proxy.
        
\subsection{Summary of proxies}

    Table \ref{table:summary} summarizes each analyzed proxy for the seismic data likelihood evaluation. It represents aspects of the computation of each proxy and the posterior results regarding the pore pressure at the validation point. The reader may note that the preeminent obstacle in using the benchmark proxy is the elapsed time to evaluate the seismic data likelihood of a single realization, even using GPU-accelerated computing. On the other hand, the proposed proxies in this work reduce the computational time by approximately 3,000-fold for the seismic data likelihood evaluation when considering a standard computer with 8 CPUs and 16 GB of memory. Nevertheless, this computational gain comes with limitations.
        
    \begin{center}
\smallskip
\begin{minipage}{0.98\linewidth}
\captionof{table}{\tcaptionOne} \label{table:summary}

\begin{tabularx}{\textwidth}{|Y| *{2}{Y} *{1}{Y|} }
\hline
\multicolumn{1}{|c|}{\bf Proxy} & \multicolumn{1}{c}{\bf Benchmark} & \multicolumn{1}{c}{\bf Semblance} & \multicolumn{1}{c|}{\bf Horizons} \\ \Xhline{6\arrayrulewidth}
Type of velocity & 
Slowness & 
RMS (time domain) &
Averaged along depth direction \\ \hline

How velocity is used &
Depth-migration of each sample &
Surface to render semblance values in velocity spectra &
Time-depth conversion of horizons \\ \hline

Metric to select posterior samples & 
Global error of $rho$ in ADCIGs ($\varepsilon_{benchmark}$) &
Inverse of mean semblance value ($\varepsilon_{semb}$) &
Computed distance on PC scores of converted horizons from original depth-positioning ($\varepsilon_{hzs}$) \\ \hline
      
Metric cut-off for ABC rejection sampling & 
$\varepsilon_{benchmark} <$ global error of legacy velocity &
$\varepsilon_{semb} <$ inverse of the mean semblance values retrieved by legacy velocity &
$\varepsilon_{hzs} <$ heuristic distance of hypothetical sample with 2\% of error from the original depth-positioning \\ \hline

Assumptions & 
Flatness of reflection events in ADCIGs represents well focused seismic images & 
Geological area with low lateral velocity variations: time migration procedures & 
Legacy seismic image in depth domain is well focused \\ \hline
                               
Seismic data use &                
Direct: Depth-migration procedures & 
Direct: Velocity spectra computation with NMO procedure & 
Indirect: Uses depth-positioning of interpreted horizons in seismic image from legacy velocity \\ \hline
          
Unmigrated seismic data manipulation &
Yes: depth-migration to generate ADCIGs & 
Yes: computing semblance velocity spectra &
No: assuming original depth-positioning of horizons embedded with seismic information \\ \hline

Elapsed time to analyze metric of one MC sample &
GPU-accelerated: approximately 50 min &
Regular computer: less than a second &
Regular computer: less than a second \\ \hline
  
Bandwidth length of pore pressure uncertainty &
Narrow &                                              
Large &                                             
Narrow \\ \hline
\end{tabularx}

\end{minipage}
\end{center}

    First, one must assume that the region of interest bears no strong lateral variation in the velocity for the semblance proxy. Moreover, a bias in selecting lower velocities is added because of the inherent hyperbolic approximation of NMO procedures in time domain. Then, the cut-off computed from the legacy velocity skews the posterior selection, yielding a significant uncertainty in the pore pressure result compared to the results of the other proxies.
    
    Second, the horizons proxy works under the assumption of valid depth-positioning of seismic horizons interpreted in a focused seismic image. Thus, the seismic data is used only indirectly for this metric computation. Despite these simplifications, horizons proxy provides similar results to the benchmark case in quantifying uncertainty for pore pressure prediction. Moreover, this proxy utilizing distances based on depth-positioning analysis carries the practicality of not dealing with unmigrated seismic data.


\section{Conclusion}

    In this paper, we discussed the implementation of an interdisciplinary methodology named Bayesian Geophysical Basin Modeling (BGBM). Furthermore, we addressed the challenge of reducing the computational demand in evaluating seismic data likelihood for Monte Carlo samples when assimilating spatial seismic data in the methodology. Besides the benchmark proxy for the seismic data likelihood, which has a high computational cost, two surrogate kinematic metrics were proposed and assessed in this work. A 2D section of a real field case in the GoM basin was considered to analyze the BGBM application with all three different proxies used to quantify uncertainty in the pore pressure prediction. The forecast of pore pressure from this methodology was validated by a mudweight measurement of a blind well near the study section.

    The two proposed proxies drastically reduce the computational cost in ${10}^3$ order of time compared to the benchmark proxy for evaluating the seismic data likelihood of Monte Carlo samples. Moreover, all pore pressure results from the different proxies could properly shift the prior distribution shape for expected pore pressure values at the validation point. Nevertheless, the semblance proxy underperformed compared to the other proxies. Even though this proxy directly uses the seismic data to build the semblance from velocity analysis, the time migration assumptions in the underlying NMO procedure introduce bias in the velocity selection of samples as posterior. Therefore, the semblance proxy is not recommended for pore pressure prediction when the structure of the basin model is built from depth-migrated seismic images.

    On the other hand, the horizons proxy produces similar results compared to the benchmark proxy. The horizons proxy uses statistical similarity analysis based on Euclidean distance in the reduced space of depth-positioning after time-depth conversion applied to seismic horizons. This surrogate model, which utilizes the velocities computed from realizations to perform time-depth conversions, has the great advantage of being computationally cheaper. Moreover, this proxy is quite convenient as no manipulation of unmigrated seismic data is required to evaluate the seismic data likelihood in the BGBM application. However, these benefits come at the cost of assumptions and limitations. The seismic stratigraphic horizons used as input in the basin modeling were assumed to be reliably interpreted on focused seismic images. Moreover, the structural positioning input was not considered as uncertain.
    
    In this particular real case application of BGBM, we highlight other aspects considered as limitations. The subsurface of the area of interest can be structurally and stratigraphically very heterogeneous; if so, one must evaluate whether time-depth conversions should use more complex procedures such as ray-tracing. Furthermore, the facies distribution in each layer must handle heterogeneity, increasing the number of uncertain parameters. In addition, deterministic calibration with optimization methods in the offset well data was still necessary to run the BGBM workflow. This calibration used the sonic log to identify moduli parameters for the mineral end-members in the rock physics model. Moreover, the constant clay content in each facies was determined using the gamma-ray log.
    
    In summation, the BGBM methodology allows the incorporation of geological expertise, well data, and seismic data based on Bayesian inference. The rock physics velocity model associated with the realizations is key to condition basin models with available seismic data. For feasibility, one must use a kinematic model to evaluate seismic data misfit for each Monte Carlo sample. Even though the workflow was used in uncertainty quantification for pore pressure prediction, BGBM can be extended to forecast and quantify the uncertainty of any related features of Monte Carlo realizations, such as porosity and vertical stress fields acquired as basin modeling outputs. BGBM workflow has enormous potential to quantify uncertainty for sub-surface properties considering the geological expertise, especially for pore pressure prediction at predrill stages. Furthermore, the increase in the BGBM efficiency – the fast BGBM presented as the horizons proxy – opens attractive avenues for uncertainty quantification when predicting sub-surface properties with seismic data calibration. One example is the consideration of different sampling strategies to enhance the posterior characterization, such as Markov chain Monte Carlo (McMC). Furthermore, a 3D basin modeling assessment of geological areas with low well data density is, in fact, feasible in the aspect of integrating the spatially exhaustive measured seismic data. However, the computational time for running 3D Monte Carlo forward basin modeling simulations for uncertainty quantification may be a bottleneck, which must be further investigated.


\section{ACKNOWLEDGEMENTS}

    We acknowledge funding from the sponsors of the Stanford Center for Earth Resources Forecasting, Petrobras, and Prof. S. Graham, Dean, School of Earth, Energy and Environmental Sciences, Stanford University. We thank A. H. Scheirer for her basin modeling guidance. We thank Schlumberger for the dataset and Petrel and PetroMod donations.

\setcounter{table}{0}
\renewcommand{\thetable}{A\arabic{table}}
\append{Details of establishing the prior distributions used in the real case application}
\label{appendix:A}

The prior geologic model of uncertainty is specified by considering the following parameters as uncertain: (1) porosity-depth constitutive model parameters capturing the mechanical compaction behavior of each lithology (ten lithologies in the field example), (2) permeability-porosity model parameters capturing the uncertainty of fluid flow for each lithology, and (3) basal heat flow boundary condition for the basin. The porosity-depth compaction model uses an \cite{Athy1930} type law. This law is shown by Equation \ref{eq:athys}.

\begin{equation}
\label{eq:athys}
    \phi = \phi_0 e^{-A_c z_e} 
\end{equation}	

Here, $\phi$ denotes the porosity of sediments, $\phi_0$ is the depositional porosity, $A_c$ denotes Athy's compaction coefficient, and $z_e$ denotes equivalent hydrostatic depth. Hydrostatic depth is the depth at which the rock having the same lithology and porosity exists under hydrostatic conditions \cite[]{BPSM_book}. The Kozeny-Carman type relation \cite[]{Ungerer1990} is used for the permeability-porosity constitutive relationship. Equation \ref{eq:phi-perm} describes this relationship.
    
\begin{eqnarray}
\label{eq:phi-perm}
    \kappa_v && = \begin{cases}
    F\frac{20\phi'^5}{S^2(1-\phi')^2} \;\;& \mbox{if~} \phi' < 10\% \\ \\
    F\frac{0.2\phi'^3}{S^2(1-\phi')^2} \;\;& \mbox{if~} \phi' > 10\% \end{cases}
\end{eqnarray}
    
Here, $\kappa_v$ represents vertical permeability, $F$ is a scaling factor, $S$ is specific surface area and $\phi' = -3.1 \times 10^{-10}S$. The horizontal permeability $\kappa_h$ is given as $a_k \times \kappa_v$, where $a_k$ denotes the permeability anisotropy factor.
    
For each of the ten lithologies in the study area, parameters $\phi_0$, $A_c$, $S$, $F$, and $a_k$ are considered uncertain. The prior probability distribution on each parameter is taken to be truncated Gaussian distributions. The mean $\mu$ and standard deviation $\sigma$ of the prior truncated Gaussian distributions are shown in Table \ref{table:priors}. The $\mu$ values were assigned based on $v_{clay}$ (Table \ref{table:priors}), which denotes the gamma-ray log-derived clay volume fraction averaged over the depth interval of each lithology. In addition, typical parameter values for pure sand and shale lithologies as specified by \cite{BPSM_book} were considered. Subsequently, mean parameter values $\mu$ were estimated using $vclay$ to interpolate between corresponding pure sand and shale values. Finally, standard deviations $\sigma$ were specified to keep the uncertainty large.

\begin{center}
\smallskip
\begin{minipage}{0.98\linewidth}
\captionof{table}{\tcaptionAOne} \label{table:priors}
\resizebox{\textwidth}{!}{%

\begin{tabular}{|c *{5}{|c} *{1}{|c|}}
\hline
\multicolumn{1}{|c}{\bf Lithology} & 
\multicolumn{1}{c}{$v_{clay}$} & 
\multicolumn{1}{c}{$A_c (km^{-1})$} & 
\multicolumn{1}{c}{$\phi_0$} & 
\multicolumn{1}{c}{$S (10^7m^{-1})$} & 
\multicolumn{1}{c}{$F$} & 
\multicolumn{1}{c|}{$a_k$} \\ \Xhline{2\arrayrulewidth}

Lithology 1 &
Unknown &
$N(0.5, 0.2)$ &  
$N(0.55, 0.06)$ & 
$N(5.05, 4)$ & 
$N(2.45, 6)$ & 
$N(3.10, 1.5)$\\ \hline

Lithology 2 &
0.60 &
$N(0.5, 0.18)$ &  
$N(0.56, 0.04)$ & 
$N(7.04, 3)$ & 
$N(1.48, 4)$ & 
$N(2.22, 1.2)$\\ \hline

Lithology 3 &
0.55 &
$N(0.44, 0.18)$ &  
$N(0.54, 0.04)$ & 
$N(6.54, 3)$ & 
$N(1.91, 4)$ & 
$N(2.41, 1.2)$\\ \hline

Lithology 4 &
0.45 &
$N(0.34, 0.18)$ &  
$N(0.51, 0.04)$ & 
$N(5.00, 3)$ & 
$N(3.15, 4)$ & 
$N(3.29, 1.2)$\\ \hline

Lithology 5 &
0.50 &
$N(0.39, 0.18)$ &  
$N(0.53, 0.04)$ & 
$N(6.05, 3)$ & 
$N(2.50, 4)$ & 
$N(2.60, 1.2)$\\ \hline

Lithology 6 &
0.36 &
$N(0.5, 0.2)$ &  
$N(0.55, 0.06)$ & 
$N(5.05, 4)$ & 
$N(2.45, 6)$ & 
$N(3.10, 1.5)$\\ \hline

Lithology 7 &
0.70 &
$N(0.55, 0.18)$ &  
$N(0.60, 0.04)$ & 
$N(8.03, 3)$ & 
$N(0.89, 4)$ & 
$N(1.84, 1.2)$\\ \hline

Lithology 8 &
0.36 &
$N(0.5, 0.2)$ &  
$N(0.55, 0.06)$ & 
$N(5.05, 4)$ & 
$N(2.45, 6)$ & 
$N(3.10, 1.5)$\\ \hline

Lithology 9 &
0.60 &
$N(0.5, 0.18)$ &  
$N(0.56, 0.04)$ & 
$N(7.04, 3)$ & 
$N(1.48, 4)$ & 
$N(2.22, 1.2)$\\ \hline

Lithology 10 &
Unknown &                                              
$N(0.5, 0.2)$ &  
$N(0.55, 0.06)$ & 
$N(5.05, 4)$ & 
$N(2.45, 6)$ & 
$N(3.10, 1.5)$\\ \hline
\end{tabular}
}
\end{minipage}
\end{center}

The gamma-ray log had missing data in the top and bottom lithologies. Consequently, larger $\sigma$ values were assigned for these lithologies. The upper and lower limits of the prior truncated Gaussian distributions are shown in Table \ref{table:limits}. The prior distribution for the basal heat flow, assumed to be spatially and temporally constant, was taken to be the Gaussian distribution with a mean and a standard deviation of 47 and 3 $mW/m^2$, respectively. Thus, a total of 51 parameters were accounted for in the prior geologic uncertainty model: Five lithological parameters for each of the ten lithologies and the basal heat flow value. The reader is referred to \cite{Pradhan2020_ABC} for additional motivations behind modeling choices and a description of other basin modeling considerations.
        
\begin{center}
\smallskip
\begin{minipage}{0.98\linewidth}
\captionsetup{justification=centering}
\captionof{table}{\tcaptionATwo} \label{table:limits}
\begin{tabularx}{\textwidth}{| Y | Y | Y | Y | Y | Y |}

\cline{2-6}
\multicolumn{1}{c}{~} & 
\multicolumn{1}{|c}{$A_c (km^{-1})$} & 
\multicolumn{1}{c}{$\phi_0$} & 
\multicolumn{1}{c}{$S (10^7m^{-1})$} & 
\multicolumn{1}{c}{$F$} & 
\multicolumn{1}{c|}{$a_k$} \\ \Xhline{4\arrayrulewidth}

Lower limit & 0.15 & 0.36 & 0.1 & 0.001 & 1.1 \\
\hline
Upper limit & 0.90 & 0.75 & 12  &    25 & 5.5 \\
\hline
\end{tabularx}
\end{minipage}
\end{center}

Given a sample from the prior distribution, basin modeling simulation is run in PetroMod software to obtain present-day porosity, pore pressure, and temperature sections. These outputs are obtained by simulating geological processes \cite[]{BPSM_book} in the basin of interest. The considered processes are layer sedimentation by lithology, mechanical compaction, fluid flow, and heat flow in the basin. In addition, we also account for chemical diagenesis in the form of the smectite-illite transition of clay minerals in shales, driven by the thermal history of the basin \cite[]{Dutta1987}. Specifically, we assume that clay minerals in the shale sediments consisted initially of 80\% smectite and 20\% illite components during deposition. This proportion is a reasonable assumption for Gulf Coast shales, as noted by \cite{Dutta1987, Dutta2016}. 

The numerical simulation uses first-order Arrhenius rate theory to model the temperature-driven smectite-illite transformation reaction. The final composition of the clay minerals, after illitization, is taken to be 20\% smectite and 80\% illite. Thus, each prior sample from the geologic prior may be related to the present-day clay mineral composition of the subsurface section.

We have assumed four constituent members for the effective medium modeling of the elastic properties of the rock: smectite, illite, nonclay stiff mineral composite for minerals such as quartz and feldspar, and brine-saturated pores. Mineral end-member elastic properties (Table \ref{table:moduli}) were calibrated using the sonic, density, and neutron-density logs available at well-1. During the calibration process, typical values for the elastic properties were randomly perturbed multiple times, and rock physics modeling was performed for each case. Because the well-1 logs were reserved to calibrate the rock-physics model, only punctual data (porosity, mudweight, and temperature) in the well was used to be further assimilated in the BGBM process.
        
\begin{center}
\smallskip
\begin{minipage}{0.98\linewidth}
\begin{center}
\captionsetup{justification=centering}
\captionof{table}{\tcaptionAThree} \label{table:moduli}
\begin{tabularx}{0.7\linewidth}{ccc}
\Xhline{3\arrayrulewidth}
              Mineral &     Parameter & Calibrated value \\
\Xhline{3\arrayrulewidth}
                    {} & Bulk modulus  &        $34.43\ \ GPa$ \\
Nonclay stiff mineral & Shear modulus &         $42.5\ \ GPa$ \\
                  {} &       Density &      $2.63\ \ g/cm^3$  \\
                  \Xhline{0.2\arrayrulewidth}
             {} & Bulk modulus  &         $11.9\ \ GPa$ \\
                  Smectite & Shear modulus &          $3.2\ \ GPa$ \\
                  {} &       Density &       $2.35\ \ g/cm^3$  \\
                  \Xhline{0.2\arrayrulewidth}
               {} & Bulk modulus  &         $39.2\ \ GPa$ \\
                  Illite & Shear modulus &        $17.3\ \ GPa$  \\
                  {} &       Density &        $2.66\ \ g/cm^3$ \\
\Xhline{3\arrayrulewidth}
\end{tabularx}
\end{center}
\end{minipage}
\end{center}

One must validate if the generated priors are statistically consistent with the available data. In other words, we must evaluate if the assumed prior distribution encompasses the "true" point retrieved from measurement. Identifying that the generated prior sub-surface properties can encompass well-1 data is easily inspected by visual means with vertical profiles plot. The falsification of priors against well data is illustrated in part (a) of Figure \ref{fig:Figs/FigureA1}.

\plot{Figs/FigureA1}{width=\columnwidth}{\captionAOne}

Moreover, to check for prior consistency against the seismic data, the MDS procedure in the resultant depth-positioning of horizons (the feature used for horizons proxy) is applied. The seismic data is represented in this space by the original depth-positioning of the horizons. The robust Mahalanobis distance (RMD) between the scores of the converted horizons of each prior sample is computed \cite[]{Rousseeuw1999}. The values of this distance using the six scores obtained in the MDS procedure, which reflect more than 99\% of prior variation, are represented in part (b) of Figure A1. The dashed line is a 95\% threshold in the RMD values. The observed data (original depth positioning of the seismic horizons after dimension reduction) falls within the dashed line and zero, indicating that the observation is not an outlier with respect to the prior distribution. Hence, the prior is not falsified compared to both data types. Consequently, one may continue with the prior in the BGBM computation to forecast properties and their uncertainty quantification.
        

\bibliographystyle{seg}         
\bibliography{BGBMpaper}

\newcommand{\SortNoop}[1]{}
\begin{thebibliography}{}
\itemsep0pt

\bibitem[Al‐Yahya, 1989]{AlYahya1989}
Al‐Yahya, K.,  1989, {Velocity analysis by iterative profile migration}:
  GEOPHYSICS, {\bf 54}, 718--729.

\bibitem[Athy, 1930]{Athy1930}
Athy, L.~F.,  1930, {Density, Porosity, and Compaction of Sedimentary Rocks1}:
  AAPG Bulletin, {\bf 14}, 1--24.

\bibitem[Avseth et~al., 2005]{Avseth2005}
Avseth, P., T. Mukerji, and G. Mavko,  2005, Quantitative seismic
  interpretation: Cambridge University Press.

\bibitem[Beaumont, 2010]{Beaumont2010}
Beaumont, M.~A.,  2010, {Approximate Bayesian computation in evolution and
  ecology}: Annual Review of Ecology, Evolution, and Systematics, {\bf 41},
  379--406.

\bibitem[Beaumont et~al., 2002]{Beaumont2002}
Beaumont, M.~A., W. Zhang, and D.~J. Balding,  2002, {Approximate Bayesian
  Computation in Population Genetics}: Genetics, {\bf 162}, 2025--2035.

\bibitem[Biondi, 2006]{Biondi2006}
Biondi, B.~L.,  2006, {3D Seismic Imaging}: Society of Exploration
  Geophysicists.

\bibitem[Blum, 2010]{Blum2010}
Blum, M. G.~B.,  2010, {Approximate Bayesian Computation: A Nonparametric
  Perspective}: Journal of the American Statistical Association, {\bf 105},
  1178--1187.

\bibitem[Brevik et~al., 2011]{Brevik2011}
Brevik, I., A. Callejon, P. Kahn, P. Janak, and D. Ebrom,  2011, {Rock
  physicists step out of the well location, meet geophysicists and geologists
  to add value in exploration analysis}: The Leading Edge, {\bf 30},
  1382--1391.

\bibitem[Brevik et~al., 2014]{Brevik_2014}
Brevik, I., T. Szydlik, M.~P. Corver, G. {De Prisco}, C. Stadtler, and H.~K.
  Helgesen,  2014, {Geophysical basin modeling: Generation of high quality
  velocity and density cubes for seismic imaging and gravity field monitoring
  in complex geology settings}: SEG Technical Program Expanded Abstracts 2014,
  Society of Exploration Geophysicists, 4733--4737.

\bibitem[Brooks et~al., 2011]{Brooks2011}
Brooks, S., A. Gelman, G. Jones, and X.-L. Meng, eds., 2011, {Handbook of
  Markov Chain Monte Carlo}: Chapman and Hall/CRC.

\bibitem[Chandler and Bate, 2007]{Chandler2007}
Chandler, R.~E., and S. Bate,  2007, {Inference for clustered data using the
  independence loglikelihood}: Biometrika, {\bf 94}, 167--183.

\bibitem[{De Prisco} et~al., 2015]{DePrisco2015}
{De Prisco}, G., D. Thanoon, R. Bachrach, I. Brevik, S.~A. Clark, M.~P. Corver,
  R.~E.~F. Pepper, T. Hantschel, H.~K. Helgesen, K. Osypov, and O.~K. Leirfall,
   2015, {Geophysical basin modeling: Effective stress, temperature, and pore
  pressure uncertainty}: Interpretation, {\bf 3}, SZ27--SZ39.

\bibitem[Doyen et~al., 2003]{Doyen2003}
Doyen, P.~M., A. Malinverno, R. Prioul, P. Hooyman, S. Noeth, L. den Boer, D.
  Psaila, C.~M. Sayers, T.~J.~H. Smit, C. van Eden, and R. Wervelman,  2003,
  {Seismic pore pressure prediction with uncertainty using a probabilistic
  mechanical earth model}: SEG Technical Program Expanded Abstracts 2003,
  Society of Exploration Geophysicists, 1366--1369.

\bibitem[Dutta, 1987]{Dutta1987}
Dutta, N.~C.,  1987, Geopressure: Society of Exploration Geophysicists.
\newblock Geophysics reprint series. No. 7.

\bibitem[Dutta, 2016]{Dutta2016}
--------, 2016, {Effect of chemical diagenesis on pore pressure in argillaceous
  sediment}: The Leading Edge, {\bf 35}, 523--527.

\bibitem[Fomel and Landa, 2014]{Fomel2014}
Fomel, S., and E. Landa,  2014, {Structural uncertainty of time-migrated
  seismic images}: Journal of Applied Geophysics, {\bf 101}, 27--30.

\bibitem[Galloway, 2008]{Galloway2008}
Galloway, W.~E.,  2008, {Chapter 15 Depositional Evolution of the Gulf of
  Mexico Sedimentary Basin}, {\it in} Sedimentary Basins of the World:
  Elsevier,  505--549.

\bibitem[Hantschel and Kauerauf, 2009]{BPSM_book}
Hantschel, T., and A.~I. Kauerauf,  2009, Fundamentals of basin and petroleum
  systems modeling: Springer Berlin Heidelberg.

\bibitem[Iversen and Tygel, 2008]{Iversen2008}
Iversen, E., and M. Tygel,  2008, {Image-ray tracing for joint 3D seismic
  velocity estimation and time-to-depth conversion}: GEOPHYSICS, {\bf 73},
  S99--S114.

\bibitem[Jiang et~al., 2017]{ABC_Jiang2017}
Jiang, B., T.~Y. Wu, C. Zheng, and W.~H. Wong,  2017, {Learning summary
  statistic for approximate Bayesian computation via deep neural network}:
  Statistica Sinica, {\bf 27}, 1595--1618.

\bibitem[Jones, 2012]{Jones2012}
Jones, I.~F.,  2012, {Tutorial: Incorporating near-surface velocity anomalies
  in pre-stack depth migration models}: First Break, {\bf 30}, 47--58.

\bibitem[Larner et~al., 1981]{Larner1981}
Larner, K.~L., L. Hatton, B.~S. Gibson, and I.~C. Hsu,  1981, {Depth migration
  of imaged time sections.}: Geophysics, {\bf 46}, 734--750.

\bibitem[Law and Spencer, 1998]{AAPGmemoir70_ch1}
Law, B., and C. Spencer,  1998, Abnormal pressures in hydrocarbon environments,
  {\it in} Abnormal pressures in hydrocarbon environments: AAPG Memoir 70.
\newblock (chapter 1, p.1-11).

\bibitem[Liu, 2004]{Book_MC_strategies}
Liu, J.~S.,  2004, {Monte Carlo Strategies in Scientific Computing}: Springer
  New York, volume~{\bf 20} {\it of} Springer Series in Statistics.

\bibitem[Mavko et~al., 2009]{book_RPH_2009}
Mavko, G., T. Mukerji, and J. Dvorkin,  2009, The rock physics handbook: Tools
  for seismic analysis of porous media, 2 ed.: Cambridge University Press.

\bibitem[Mukerji et~al., 2002]{Mukerji2002_CSEG}
Mukerji, T., N. Dutta, M. Prasad, and J. Dvorkin,  2002, {Seismic Detection and
  Estimation of Overpressures Part I: the Rock Physics Basis}: CSEG Recorder,
  {\bf 27}, 34--57.

\bibitem[Osborne and Swarbrick, 1997]{Osborne_AAPG97}
Osborne, M.~J., and R.~E. Swarbrick,  1997, {Mechanisms for Generating
  Overpressure in Sedimentary Basins: A Reevaluation}: AAPG Bulletin, {\bf 81},
  1023--1041.

\bibitem[Park et~al., 2013]{Park2013}
Park, H., C. Scheidt, D. Fenwick, A. Boucher, and J. Caers,  2013, {History
  matching and uncertainty quantification of facies models with multiple
  geological interpretations}: Computational Geosciences, {\bf 17}, 609--621.

\bibitem[Peng et~al., 2007]{Peng2007}
Peng, C.~C., A. Yu, J. Ratcliffe, and C. Harding,  2007, {3D basin‐modeling
  approach to subsalt prestack depth imaging: An example from the Atwater fold
  belt, deepwater of GOM}: SEG Technical Program Expanded Abstracts 2007,
  Society of Exploration Geophysicists, 402--406.

\bibitem[Petmecky et~al., 2009]{Petmecky2009}
Petmecky, R.~S., M.~L. Albertin, and N. Burke,  2009, {Improving sub-salt
  imaging using 3D basin model derived velocities}: Marine and Petroleum
  Geology, {\bf 26}, 457--463.

\bibitem[Pon and {R. Lines}, 2005]{Pon2005}
Pon, S., and L. {R. Lines},  2005, {Sensitivity analysis of seismic depth
  migrations}: GEOPHYSICS, {\bf 70}, S39--S42.

\bibitem[Pradhan et~al., 2020]{Pradhan2020_ABC}
Pradhan, A., N.~C. Dutta, H.~Q. Le, B. Biondi, and T. Mukerji,  2020,
  {Approximate Bayesian inference of seismic velocity and pore-pressure
  uncertainty with basin modeling, rock physics, and imaging constraints}:
  Geophysics, {\bf 85}, ID19--ID34.

\bibitem[Pradhan and Mukerji, 2020]{Pradhan2020_SBEL}
Pradhan, A., and T. Mukerji,  2020, {Seismic Bayesian evidential learning:
  estimation and uncertainty quantification of sub-resolution reservoir
  properties}: Computational Geosciences, {\bf 24}, 1121--1140.

\bibitem[Rousseeuw and Driessen, 1999]{Rousseeuw1999}
Rousseeuw, P., and K. Driessen,  1999, {A Fast Algorithm for the Minimum
  Covariance}: Technometrics, {\bf 41}, 212--223.

\bibitem[Sava and Fomel, 2003]{Sava2003}
Sava, P.~C., and S. Fomel,  2003, {Angle-domain common-image gathers by
  wavefield continuation methods}: Geophysics, {\bf 68}, 1065--1074.

\bibitem[Sayers et~al., 2002]{Sayers02}
Sayers, C.~M., G.~M. Johnson, and G. Denyer,  2002, Predrill pore-pressure
  prediction using seismic data: Geophysics, {\bf 67}, 1286--1292.

\bibitem[Scheidt et~al., 2015]{Scheidt2015}
Scheidt, C., C. Jeong, T. Mukerji, and J. Caers,  2015, {Probabilistic
  falsification of prior geologic uncertainty with seismic amplitude data:
  Application to a turbidite reservoir case}: Geophysics, {\bf 80}, M89--M100.

\bibitem[Scheidt et~al., 2018]{QUSS_book}
Scheidt, C., L. Li, and J. Caers,  2018, {Quantifying Uncertainty in Subsurface
  Systems}: John Wiley {\&} Sons, Inc., volume~{\bf 29} {\it of} Geophysical
  Monograph Series.

\bibitem[Sisson et~al., 2018]{Handbook_of_ABC}
Sisson, S., Y. Fan, and M. Beaumont,  2018, {Handbook of Approximate Bayesian
  Computation}, 1st ed.: Chapman and Hall/CRC.

\bibitem[Szydlik et~al., 2015]{Szydlik2015}
Szydlik, T., H.~K. Helgesen, I. Brevik, G. {De Prisco}, S.~A. Clark, O.~K.
  Leirfall, K. Duffaut, C. Stadtler, and M. Cogan,  2015, {Geophysical basin
  modeling: Methodology and application in deepwater Gulf of Mexico}:
  Interpretation, {\bf 3}, SZ49--SZ58.

\bibitem[Tarantola, 2006]{Tarantola2006}
Tarantola, A.,  2006, {Popper, Bayes and the inverse problem}: Nature Physics,
  {\bf 2}, 492--494.

\bibitem[Ungerer et~al., 1990]{Ungerer1990}
Ungerer, P., J. Burrus, B. Doligez, P.~Y. Chenet, and F. Bessis,  1990, {Basin
  evaluation by integrated two-dimensional modeling of heat transfer, fluid
  flow, hydrocarbon generation, and migration}.

\bibitem[Varin et~al., 2011]{Varin2011}
Varin, C., N. Reid, and D. Firth,  2011, {An overview of composite likelihood
  methods}: Statistica Sinica, {\bf 21}, 5--42.

\bibitem[Virieux and Operto, 2009]{Virieux2009}
Virieux, J., and S. Operto,  2009, {An overview of FWI in Exploration
  Geophysics}: Geophysics, {\bf 74}, WCC127--WCC152.

\bibitem[Wang, 2012]{Wang2012}
Wang, J.,  2012, 6, {\it in} Classical Multidimensional Scaling: Springer
  Berlin Heidelberg,  115--129.

\bibitem[Yilmaz, 2001]{Yilmaz2001}
Yilmaz, {\"{O}}.,  2001, {Seismic Data Analysis}: Society of Exploration
  Geophysicists.

\end{thebibliography}

\end{document}